\def\ZPC#1#2#3{{\sl Z.~Phys.} {\bf C#1}~(#3) #2}

\def\PRL#1#2#3{{\sl Phys. Rev. Lett.} {\bf #1}~(#3) #2}
\def\PRD#1#2#3{{\sl Phys. Rev.} {\bf D#1}~(#3) #2}
\def\PLB#1#2#3{{\sl Phys. Lett.} {\bf B#1}~(#3) #2}
\def\PREP#1#2#3{{\sl Phys. Rep.} {\bf #1}~(#3) #2}
\def\NPB#1#2#3{{\sl Nucl. Phys.} {\bf B#1}~(#3) #2}

\def\beq{\begin{equation}}
\def\eeq{\end{equation}}
\def\bea{\begin{eqnarray}}
\def\eea{\end{eqnarray}}
\def\bq{\begin{quote}}
\def\eq{\end{quote}}

\def\gappeq{\mathrel{\rlap {\raise.5ex\hbox{$>$}}
{\lower.5ex\hbox{$\sim$}}}}

\def\lappeq{\mathrel{\rlap{\raise.5ex\hbox{$<$}}
{\lower.5ex\hbox{$\sim$}}}}

\def\bbz{fa Z \kern-8.9pt Z}
\documentstyle [12pt,epsfig]{article}
  \newcommand{\ccaption}[2]{
    \begin{center}
    \parbox{0.85\textwidth}{
      \caption[#1]{\small{\it{#2}}}
      }
    \end{center}
    }
\begin{document}
\thispagestyle{empty}
\vspace*{-1cm}
\begin{flushright}
{CERN-TH/97-40} \\
{hep-ph/9703276} \\
\end{flushright}
\vspace{1cm}
\begin{center}
{\large {\bf Pursuing Interpretations of the HERA Large-Q$^2$ Data}} \\
\vspace{.2cm}                                                  
\end{center}

\begin{center}
\vspace{.3cm}
{\bf G. Altarelli\footnote{Also at Universit\`a di Roma III, Rome, Italy.}, 
J. Ellis,                                                                
G.F. Giudice\footnote{On leave of absence from INFN, Sez. di Padova, Italy.}, 
S. Lola} and                                                     
{\bf M.L. Mangano}\footnote{On leave of absence from INFN, 
                            Sez. di Pisa, Italy.} 
\\                          
\vspace{.5cm}
{ Theory Division, CERN, CH-1211, Gen\`eve 23, Switzerland} \\
\end{center}                              

\vspace{2cm}

\begin{abstract}
We explore interpretations of the anomaly
observed by H1 and ZEUS at HERA in deep-inelastic $e^+p$ scattering at
very large $Q^2$, in terms of possible physics beyond the Standard Model.
Since the present data could be compatible with either a continuum or a
resonant solution, we
discuss both the possibilities of new effective interactions and the
production of a narrow
state of mass $M\sim 200$ GeV with leptoquark couplings. We compare
these models with the measured $Q^2$ distributions:
for the contact terms, constraints from LEP~2 and the
Tevatron allow only a few choices of helicity and flavour structure
that could roughly fit the HERA data. The data are instead quite consistent with
the $Q^2$ distribution expected from a
leptoquark state. We study the production cross sections of such a
particle at the Tevatron and at HERA, the latter in the cases where it is
produced from either a valence or a sea quark. The absence of a signal at
the
Tevatron disfavours the likelihood that any such leptoquark decays only
into $e^+ q$. We then
focus on the possibility that the leptoquark is a squark with $R$-violating 
couplings. In view of the present experimental limits on such 
couplings, the most likely production channels are
$e^+d\rightarrow \tilde c_L$ or perhaps $e^+d\rightarrow \tilde t$, with
$e^+s\rightarrow
\tilde t$ a more marginal possibility. We point out that the $\tilde c_L$
could have competing branching ratios for
$R$-conserving and $R$-violating decay channels, whereas $\tilde t$ decays
would be more likely to be dominated by one or the other.
Possible tests of our preferred model include the absence both of
analogous events in $e^- p$ collisions and of charged current
events, and the presence of detectable cascade decays whose
kinematical signatures we discuss. This model could also make an
observable contribution to $K \rightarrow \pi {\bar \nu} \nu$ and/or
neutrinoless $\beta\beta$ decay. We also discuss the possible implications for
the Tevatron and for $e^+ e^- \rightarrow {\bar q} q$ and neutralinos at LEP~2.
\end{abstract}
\vfill 
\begin{flushleft}
CERN-TH/97-40\\
March 1997\\
\end{flushleft}

\newpage

\section{Introduction and Summary}
The HERA experiments H1~\cite{H1} and ZEUS~\cite{ZEUS} have recently
reported an excess of deep-inelastic $e^+p$
scattering events at large values of $Q^2>1.5 \times 10^4$ GeV$^2$, in a
domain not previously explored by other
experiments. With a total $e^+p$ integrated luminosity of 14 pb$^{-1}$, 
H1~\cite{H1} observes 7 events with
large $e^+$-jet invariant masses $M=\sqrt{xs}$, clustered around $M=200$
GeV, in which the positron
is backscattered at large $y=Q^2/M^2$. Similarly, ZEUS~\cite{ZEUS} with an
integrated luminosity of
20 pb$^{-1}$ observes 5 events at comparable large values of $Q^2$, $x$ 
and $y$. Although the H1 and
ZEUS data are mutually consistent and the presence of the same type of
excess in the two
experiments is certainly impressive, the detailed features of the events
are not exactly the same
in H1 and ZEUS. The events of H1 are more suggestive of a resonance with
$e^+$-quark quantum
numbers than the ZEUS data points, which are more scattered in mass. The
difference could, however,
be due to the different methods of mass reconstruction used by the two
experiments, or to fluctuations in the event characteristics. Of course, 
at this stage, due to the limited
statistics, one cannot exclude the possibility that the whole effect is
a statistical fluctuation. This will hopefully 
be clarified soon by the coming 1997 run. Meanwhile, it is important to
explore possible
interpretations of the signal, in particular with the aim of identifying
additional
signatures that might eventually be able to discriminate between different
explanations of the reported excess.

Since the observed excess is with respect to the Standard Model
expectation based on the QCD-improved parton model, the first question is 
whether the effect could be explained by some 
inadequacy of the conventional analysis without invoking new physics
beyond the Standard Model. In the 
case of the apparent excess of jet production at large
transverse energy $E_T$
recently observed by the CDF collaboration at the Tevatron~\cite{CDF}, it
has been argued~\cite{CTEQ} that a substantial
decrease in the discrepancy can be obtained by modifying the gluon parton
density at large values
of $x$ where it has not been measured directly. In the HERA
case~\cite{H1,ZEUS}, a similar explanation is apparently not
viable. In this case, the valence quark densities are the most relevant
ones, and they have been measured 
directly~\cite{BCDMS} in the same range of $x$ at much lower
values of $Q^2$. Since the values of $x$ which are relevant for the HERA excess
are quite large ($x\sim 0.5$), it is possible in principle that higher-order
effects of the Sudakov type, not accounted for by the standard
next-to-leading-order QCD analysis of the structure function data, could affect
the low-energy extraction of the partonic densities and their evolution to high
$Q^2$~\cite{Catani96}.
It should be remarked, however, that the most recent measurements of $F_2$
from
the HERA experiments explore the same high-$x$ values in a
large range of $Q^2$ values, and no anomaly was found at large $x$ up to
$Q^2\sim 10^4$ GeV$^2$~\cite{H1,ZEUS}. The evolution
logarithms could not explain the abrupt occurrence of the
effect,
which is undetected at 
$Q^2\sim 10^4$ GeV$^2$ but fully visible at $Q^2\sim 2 \times 10^4$
GeV$^2$. Therefore one can safely conclude that Sudakov effects cannot provide
a credible explanation of the observed excess.
We also note that, if the parton densities
were to blame, very similar effects should be seen in both neutral and
charged current channels, with both
$e^+$ and $e^-$ beams. This can be checked in the near future. We
do not consider this
alternative in the following, but concentrate on interpretations
based on possible new physics.

We first discuss the possibility that the observed excess is a
non-resonant continuum. Within this scenario, a
rather general approach is to interpret the HERA excess as due to an
effective four-fermion ${\bar e} e {\bar q} q$ contact
interaction~\cite{Eichten83}
with a scale $\Lambda$ of order 1.5-2.5 TeV. It is
interesting that a similar contact term of the
${\bar q} q {\bar q} q$ type, with a scale of exactly the same order of
magnitude, could also reproduce
the CDF excess in jet production at large $E_T$~\footnote{Note,
however, that this interpretation is not strengthened by 
more recent data on
the dijet angular distribution~\cite{CDFangle}.}. We study
the contact interaction scenario for the HERA excess in
some detail. In
order to interfere with the electroweak gauge interactions, the contact
term is taken as
the local product of two mixed vector- and axial-vector currents. We study
the $x$ and
$Q^2$ distributions that correspond to different flavours, signs and
choices of helicity for the contact terms (i.e.,
$LL, RR, LR$ or $RL$), and compare them with the HERA data.
Strong bounds on the possible magnitudes of the interaction
scales $\Lambda$ are imposed by CDF data~\cite{Bodek96} on the
Drell-Yan production of $e^+e^-$ pairs at large invariant mass, and by
LEP~2 data on hadron production~\cite{Opal96,Opal97}. 
If we restrict our analysis to one particular term at a time, though in
general
one cannot exclude a superposition of different chiral structures, we find
that most of the individual contact terms that could fit the HERA
data are already
excluded.  Only for particular choices of quark flavour, sign and
helicity can one obtain even rough agreement with
the HERA data while escaping the existing bounds. We present examples of
these models, and point out
the desirability of further tests at LEP~2, where a complete analysis
by all the experimental collaborations is still lacking, and at
the Tevatron. It is interesting that the existence of the appropriate
contact terms could be soon
excluded, or their effects discovered in these experiments. 
We recall that the effects of contact terms should
be present in both the $e^+$ and the $e^-$ cases with the same intensity,
and possibly also in the charged current channel, if left-handed currents are
invoked.     

We then focus on the possibility of a resonance with $e^+ q$ quantum
numbers, namely a leptoquark. Most
probably the production at HERA occurs from valence $u$ or $d$ quarks,
since otherwise the coupling would need to be
quite large, and more difficult to reconcile with existing
limits~\cite{Buch86,lqlimits,Opal97,H1limit}. Assuming an $S$-wave state, one may
have either a scalar or a vector leptoquark. Although we mostly consider
the first option, we also include some discussion of the vector
case. Defining the coupling $\lambda$ for a scalar $\phi$ by
$\lambda \phi {\bar e}_L
q_R$ or $\lambda \phi {\bar e}_R q_L$,
the observed excess of $\sim 10$ events in 34 pb$^{-1}$, 
observed with an efficiency of $\sim80\%$, suggests values of
$\lambda \sim 0.025$ or $0.04$ for production from $u$ or $d$ quarks
respectively. The corresponding natural decay width
is of the order of a few MeV. 
We compute the $Q^2$ distribution predicted by
leptoquark production, and show that it matches
the data better than the corresponding distributions for
contact terms. 
A scalar $e^+u$ or $e^+d$ state couples
to the following SU(2)
doublet combinations: $e^+_L (u_L, d_L)$, $(e^+_R, \bar\nu_R)u_R$, 
or $(e^+_R, \bar\nu_R)d_R$.
This implies that a scalar with $e^+u$ or $e^+d$ 
quantum numbers can decay into $\bar\nu q$ final
states only if it has another Yukawa interaction besides
that responsible for its production. 
This additional interaction involves a lepton field of opposite chirality
and is strongly constrained by pion decays.
In its absence, a leptoquark
would not be able to explain any resonant signal
in the charged-current channel. The H1 collaboration has
reported~\cite{H1} four
events in this channel, with a Standard Model background of about two,
but ZEUS~\cite{ZEUS} has not reported a recent charged-current
analysis. We note that the situation is different for a
vector leptoquark in the $e^+q$ case, or for a scalar leptoquark in the
$e^-q$ case. In the latter
case we could, for example, have a coupling to the weak isospin singlet
$e^-_Lu_L-\nu_Ld_L$ that indeed leads to both
neutral and charged current decay modes.

Leptoquarks would be produced via QCD interactions at the
Tevatron~\cite{TevLQ}. 
We find that a scalar leptoquark of mass $M \sim 200$ GeV has a
production cross section of
around 0.2 pb at the Tevatron. The cross section for vector leptoquarks is
somewhat model-dependent, but expected to be much larger, as
discussed later. Given the large value of the cross section, a
leptoquark branching ratio ${\cal B}(e^+ q) < 1$
into the observed $e$-jet channel is perhaps needed, even in
the scalar case, to avoid a possible combined CDF/D0 exclusion limit.
The current best limit from the Tevatron for a first-generation leptoquark is
194~GeV for ${\cal B}(e^+ q) = 1$,
recently given by the D0 collaboration~\cite{D0LQ}, and the corresponding
limit for 
${\cal B}(e^+ q) = {\cal B}(\bar\nu q) =0.5$ is 143~GeV~\cite{D0LQ}.
Thus any scalar with leptoquark quantum numbers might need additional
decay modes beyond those given by the $\lambda$ interaction introduced
above for its production mechanism.

Perhaps the most appealing form of leptoquark is a squark \cite{MSSM} with 
couplings that violate $R$ parity~\cite{RPth}. This
possibility has been put forward in connection with the HERA events also
in ref.~\cite{CR}. In terms of
supersymmetric chiral multiplets, the relevant coupling is given by
\begin{equation}
\lambda'_{ijk}L_iQ_jD^c_k
\label{1}
\end{equation}
where $L_i$, $Q_j$ and $D^c_k$ are superfields of lepton doublets, quark
doublets and quark
singlets respectively, and $i,j,k$ are generation indices. Leaving
production from the sea aside for the moment, the processes relevant for
HERA that arise from this coupling are
$e^+_R d_R \rightarrow {\tilde u}_L$ or ${\tilde c}_L$ or ${\tilde t}_L$.
We find that the
first possibility is eliminated by existing limits on $\lambda'_{111}$,
in particular those from $\beta\beta$ decay~\cite{Hirsch},
whereas the latter are still permitted. In
the following, we study
the scharm $\tilde c$ and stop $\tilde t$ possibilities in 
some detail. 
We recall that the
$R$-violating decays of the produced squark must
compete with the ordinary $R$-conserving decays.
Although some sizeable additional decay channels are welcome in view of
the
non-observation of a signal at CDF/D0, the $R$-conserving channels
should have a moderate rate, otherwise the
coupling $\lambda'_{ijk}$ required to explain the HERA excess
becomes dangerously large, particularly in view of upper limits
on $\lambda'_{121}$ coming from $\beta\beta$~\cite{Hirsch} 
and $K \rightarrow \pi {\bar \nu} \nu$~\cite{Agashe} decays. We make
a careful study of the regions of parameter space for the
supersymmetric model where the balance of $R$-conserving and $R$-violating
decays is favourable, for both
the $\tilde c$ and $\tilde t$ cases. Such a balance is more likely in
the ${\tilde c}_L$ case than for the ${\tilde t}$. In the squark scenario
there would be no signal in the
charged-current ${\bar \nu} q$ channel, but there could be $R$-conserving
decay signatures, whose kinematical properties we discuss
later. No signal is expected from the same sparticle with $e^-$ beams,
unless one is sensitive to
production from the ${\bar d}$ sea density. A distinctive signature of
the $\tilde c$ possibility
could be the appearance of a signal in $K \rightarrow \pi {\bar \nu} \nu$
close to the present upper
limit. 

We have also examined possible
signatures of $R$-violating supersymmetric interactions at LEP~2. 
We find that interference effects
in the reactions $e^+e^-\rightarrow {\bar q} q$ are unlikely to be
detectable, unless the HERA squark is produced from the sea, in
which case a signal might be detectable in ${\bar s} s$ final states. 
However, other effects of
$R$-parity violation could be observable, such as $e^+e^-\rightarrow
\chi^0 \chi^0$, where $\chi^0$ denotes the lightest neutralino,
thanks to its $R$-violating decays. 
We also discuss the compatibility of the squark
explanation of the HERA events with the model recently
proposed~\cite{CGLW} to interpret the four-jet anomaly
found by ALEPH~\cite{ALEPH}, but not seen by the other LEP
experiments~\cite{Schlatter}.

\section{Effective Contact Interactions}
Whereas the H1 data are at first sight quite suggestive of the production
of a
resonance in the $s$ channel, the spread in $x_{2\alpha}$ of the ZEUS
data~\footnote{The variable $x_{2\alpha}$ is the Bjorken $x$ variable extracted
using the double-angle  method~\cite{H1,ZEUS}.
We discuss in the Appendix issues related to this spread  and to the comparison
between values of $x$ reported by H1~\cite{H1} and ZEUS~\cite{ZEUS}, in the
light of possible initial-state radiation.}
seem to favour the possibility of an effective 4-fermion contact
interaction, which we pursue first as a more conservative option. 
As is customary in the literature~\cite{Eichten83}, we parametrize the
contact
interactions in terms of the mass scale $\Lambda$ appearing 
in the following effective Lagrangian:
\begin{equation}  \label{eq:compo}            
    {\cal L}_4 \; = \; 4\pi \, 
    \sum_{\scriptstyle i,j=L,R \atop \scriptstyle q=u,d} \, 
    \frac{\eta_{ij}}{(\Lambda^q_{ij})^2}
     \, \bar{e}_i \gamma^{\mu} e_i \, \bar{q}_j \gamma_{\mu} q_j \; .
\end{equation}
We allow for independent couplings of $u$ and $d$ quarks, as well as for
independent couplings of all different helicity states~\footnote{We do
not consider here scalar current couplings, because they are very
strongly constrained by low-energy data on helicity-suppressed
decays~\cite{Buch86,lqlimits}.}.
The parameter $\eta_{ij}$ takes the values $\pm 1$, and allows for constructive
and destructive interferences in the different channels.
                                                        
Very tight constraints on the size of such possible interactions have been set
in the past~\cite{Barnett96}. Recent preliminary results from
dielectron production at the Tevatron~\cite{Bodek96} and from hadron
production at LEP~2~\cite{Opal96,Opal97}
restrict even further the allowed ranges of the parameters
$\Lambda^q_{ij}$ in (\ref{eq:compo}). 
The most recent analysis by OPAL~\cite{Opal97}, in particular, sets
the 95\% CL limits on
the 16 independent couplings in (\ref{eq:compo}) shown in
Table~\ref{tab:lambda}. 
\begin{table}
\begin{center}
\begin{tabular}{|l|cc|cc|} \hline
 $q$   &                 
    \multicolumn{2}{c|}{$e^+e^- \to u \bar{u}$} & 
    \multicolumn{2}{c|}{$e^+e^- \to d \bar{d}$} \\
 $ij$ &  $\eta=+1$ &  $\eta=-1$ &  $\eta=+1$ &  $\eta=-1$ \\
\hline
 LL & 1.1 & 2.4 & 2.4 & 1.0 \\              
 RR & 1.4 & 1.7 & 2.1 & 1.2 \\
 LR & 1.5 & 1.6 & 1.7 & 1.4 \\
 RL & 1.7 & 1.4 & 1.6 & 1.5 \\ \hline
\end{tabular}                                     
\ccaption{}{\label{tab:lambda}
Preliminary OPAL 95\% CL limits~\cite{Opal97}
on the effective contact interaction scale $\Lambda^q_{ij}$ (in TeV). }                  
\end{center}                                                                
\end{table} 
The CDF limits~\cite{Bodek96} have only been given for the isoscalar
$u+d$ quark flavour combination, and for the $LL$ helicity combination.
We have simulated the effects in hadronic collisions of the
contact interactions for which the OPAL limits allow good fits
of the HERA data. For these cases
we extracted from the Tevatron dilepton data limits on the relevant
parameters $\Lambda^q_{ij}$
by analogy with the limits provided by CDF for the isoscalar $LL$
case~\cite{Bodek96}. 
                                                                             
Since contact interactions do not generate any particular structure
in the $x$ distributions, we discuss their impact
on the integrated $Q^2$ distributions of
the HERA data, as provided
in their papers and combined in their public presentations~\cite{H1,ZEUS}.
We have calculated the effects of each one of the 16 4-fermion couplings,
including its interference with the Standard Model DIS processes. In the
cases where the OPAL constraint allows a fit, we have applied the
inferred CDF constraints on the corresponding
$\Lambda^q_{ij}$, which are usually stronger.

\begin{figure}             
\centerline{\epsfig{figure=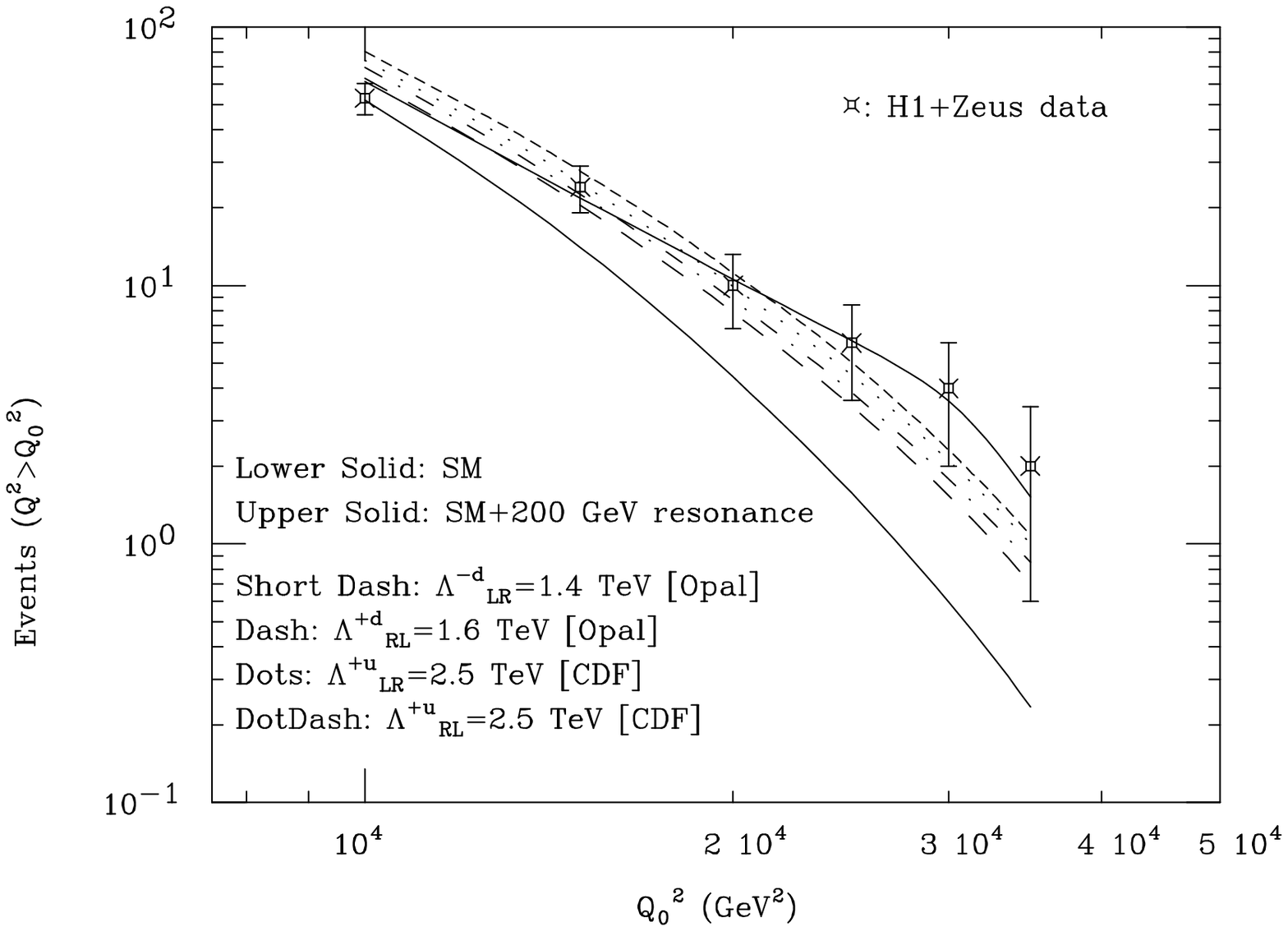,width=0.85\textwidth,clip=}}
\ccaption{}{ \label{fig:complim}    
HERA data for the integrated $Q^2$ distribution, compared to current limits on
effective contact interactions from OPAL and CDF. Only the four
combinations that best reproduce
the HERA data are shown. The lower solid line corresponds to
the prediction of the Standard Model. The upper solid curve corresponds to
the 
decay of a 200~GeV $s$-channel resonance, produced at a rate compatible
with the reported excess of events. }
\end{figure}

We applied the analysis                     
cuts of the H1 and ZEUS experiments, as described in their publications, and
combined the expectations for the respective integrated luminosities and
efficiencies~\footnote{We have verified that our Standard Model 
predictions, after accounting
for the analysis cuts, efficiencies and integrated luminosities of the two
experiments, agree with those presented in the H1 and ZEUS
papers~\cite{H1,ZEUS}.}. 
Only a few of the 16 possible couplings are at all compatible
with the HERA
data, and the four best cases are presented in Fig.~\ref{fig:complim}. In
none of
these cases is the agreement in shape between data and expectations
particularly good.
We note the essential r\^oles played by both OPAL and CDF in
constraining the possible effective contact interactions: 
in particular, there
are good fits of the HERA data for couplings with magnitudes that
are at best compatible with the OPAL data alone,
such as the choice $\Lambda^u_{RL}=1.4$ TeV, with $\eta=-1$.
However, this possibility is excluded by the
CDF limit, which is stronger for this specific coupling, namely
larger than 3~TeV.

We conclude from this study 
that, while the contact interaction hypothesis cannot be
entirely ruled out, the strong constraints already set by the LEP and Tevatron
experiments do not allow for good fits of either the event rates or the $Q^2$
distributions of the H1 and ZEUS data. We remark once more that, in any
case,
for the contact interaction hypothesis to be tenable, the 
apparent resonant structure in the invariant $e^+ q$ mass distribution
reported by H1 could only be the result of a statistical fluctuation, 
and should be washed out by higher statistics.
We would also expect that 
a joint effort of the four LEP collaborations using the combined
set of all LEP~2 data already on tape could further restrict the allowed
ranges of the $\Lambda$ parameters, as could a combined analysis of
CDF and D0 data.

\section{Leptoquarks}
Since the excess found by
H1~\cite{H1} occurs in a small range of $e^+ q$ invariant
masses~\footnote{Although the ZEUS events are more spread in
invariant mass, and appear at larger values of $x_{2\alpha}$,
we note that an inter-collaboration working group has found
that the two data sets are compatible, and that ISR 
and other instrumental effects could in
principle cause shifts of several \% in the observed
values of $x_{2\alpha}$, as discussed in the Appendix.}, it is
natural to examine models containing a new boson
with leptoquark quantum numbers,
which may be classified
according to their spin and isospin quantum numbers~\cite{Buchmueller87}.
As discussed in the introduction, they may couple      
to a fermionic current constructed out of a quark and a lepton with a
coupling
constant $\lambda$. In the narrow-width approximation,
the leading-order parton-level cross section for
production of a leptoquark of spin $J$ at
HERA is given by:                                            
\beq    \label{eq:leptoquark}
      \sigma \; = \; \frac{\pi}{4 s} \, \lambda^2 \, F_J
  \quad \quad (F_0=1, \; F_1=2) \; .
\eeq
\begin{figure}
\centerline{\epsfig{figure=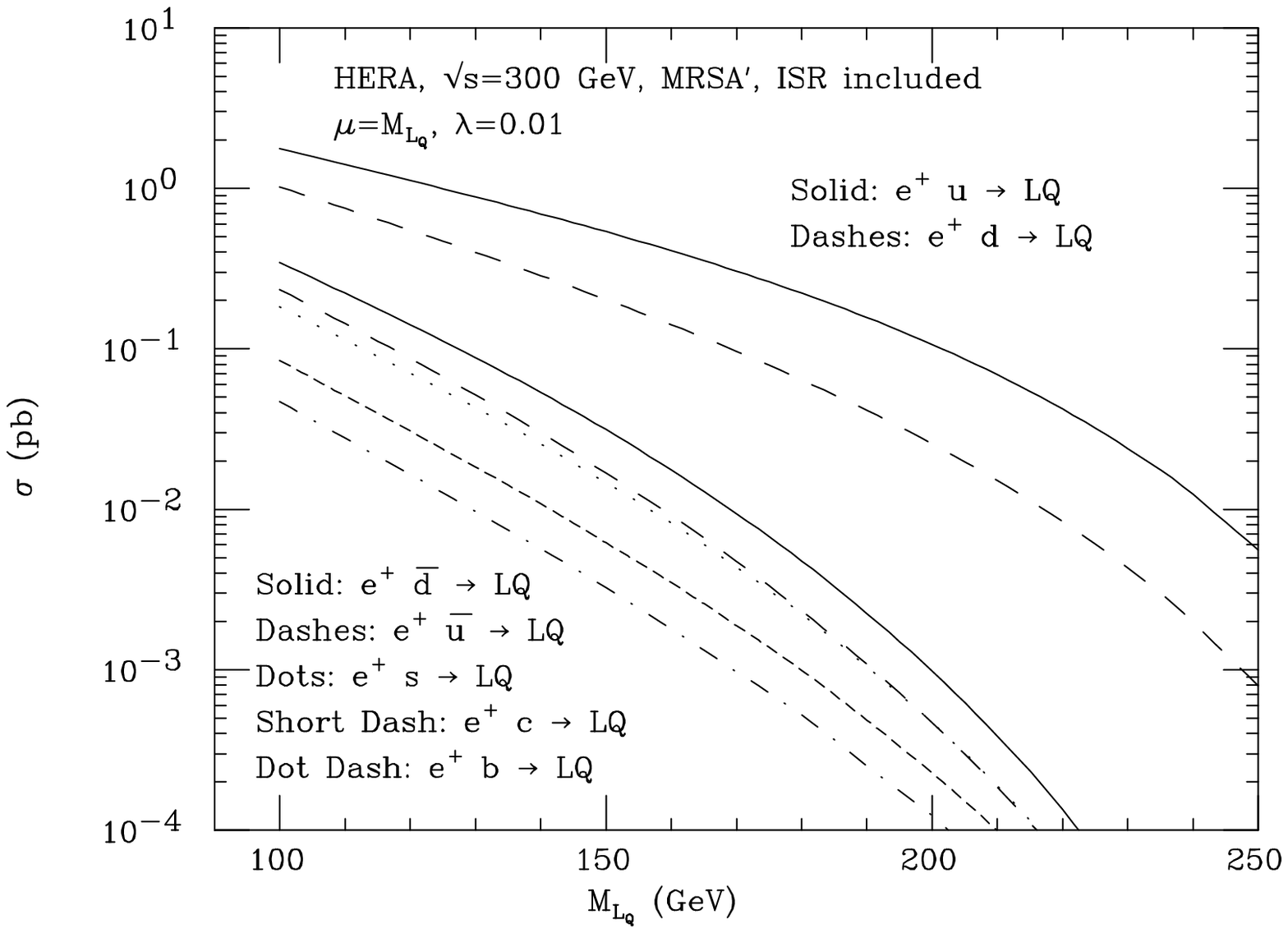,width=0.85\textwidth,clip=}}
\ccaption{}{ \label{fig:lqxsec}    
Scalar leptoquark (or $R$-violating squark) production cross sections at
HERA, including the
effects of initial-state radiation (ISR). The contributions of different
quark flavours in the proton are shown
separately.}          
\end{figure}
The convolution of the parton cross section (\ref{eq:leptoquark})
with the parton densities of the proton~\cite{Martin95}
and with the effects of
initial-state photon radiation from the positron, yields the 
cross sections shown in Fig.~\ref{fig:lqxsec}. We have assumed 
in this figure a leptoquark
coupling to only one fermionic helicity, and have used $\lambda=0.01$ as
a reference value.
Figure~\ref{fig:lqxsec} shows results for all quark flavours.
As is clear from the figure, all effective sea quark
luminosities are
significantly suppressed for masses in the 200~GeV region, where the
hypothetical HERA signal lies. 
The cross sections for $\lambda = 0.01$ in the different production
channels to produce a leptoquark with mass 200~GeV
are given in Table~\ref{tab:lqxsec}.
{
\renewcommand{\baselinestretch}{1.5}
\begin{table}
\begin{center}
\begin{tabular}{|l|ccccccc|} \hline
$\sigma$(fb), $\lambda=0.01$    
      & $e^+ u$ & $e^+ d$ & $e^+ \bar{d}$ & $e^+ \bar{u}$ & $e^+ s$ & $e^+ c$ 
& $e^+ b$ \\ \hline                     
(with ISR)& 106 & 25.6 & 0.98 & 0.47 & 0.47 & 0.23 & 0.12 \\
(no ISR)  & 117 & 28.4 & 1.12 & 0.53 & 0.53 & 0.26 & 0.14 \\  \hline
\end{tabular}                                     
\ccaption{}{\label{tab:lqxsec}
Production cross sections for a scalar leptoquark (or $R$-violating 
squark) of mass 200~GeV
at HERA, assuming a coupling $\lambda = 0.01$, showing the effects of
ISR.}
\end{center}                                                                   
\end{table} 
}
We have verified that the acceptance cuts imposed by H1 and ZEUS on their
data have
efficiencies of approximately 80\% and 100\%,
respectively. Accounting for the detector
efficiencies of the two experiments as quoted in their papers (of the order of
80\%), for the relevant integrated luminosities, and assuming a total of
10 signal events in the combined experiments, we find that a value of
$\lambda$ of approximately $0.04/\sqrt{\cal B}$ is required for leptoquark
production by $e^+d$ collisions, with correspondingly larger
couplings required for production by $e^+$ collisions with sea quarks. The
cross sections for a vector leptoquark are a factor of two larger, and 
hence would require couplings smaller by a factor of $\sqrt{2}$.
The implications of the value $\lambda \sim 0.04$ in the case of scalar
quark production via an $R$-violating interaction will be discussed
in the following section.                                                    

\begin{figure}
\centerline{\epsfig{figure=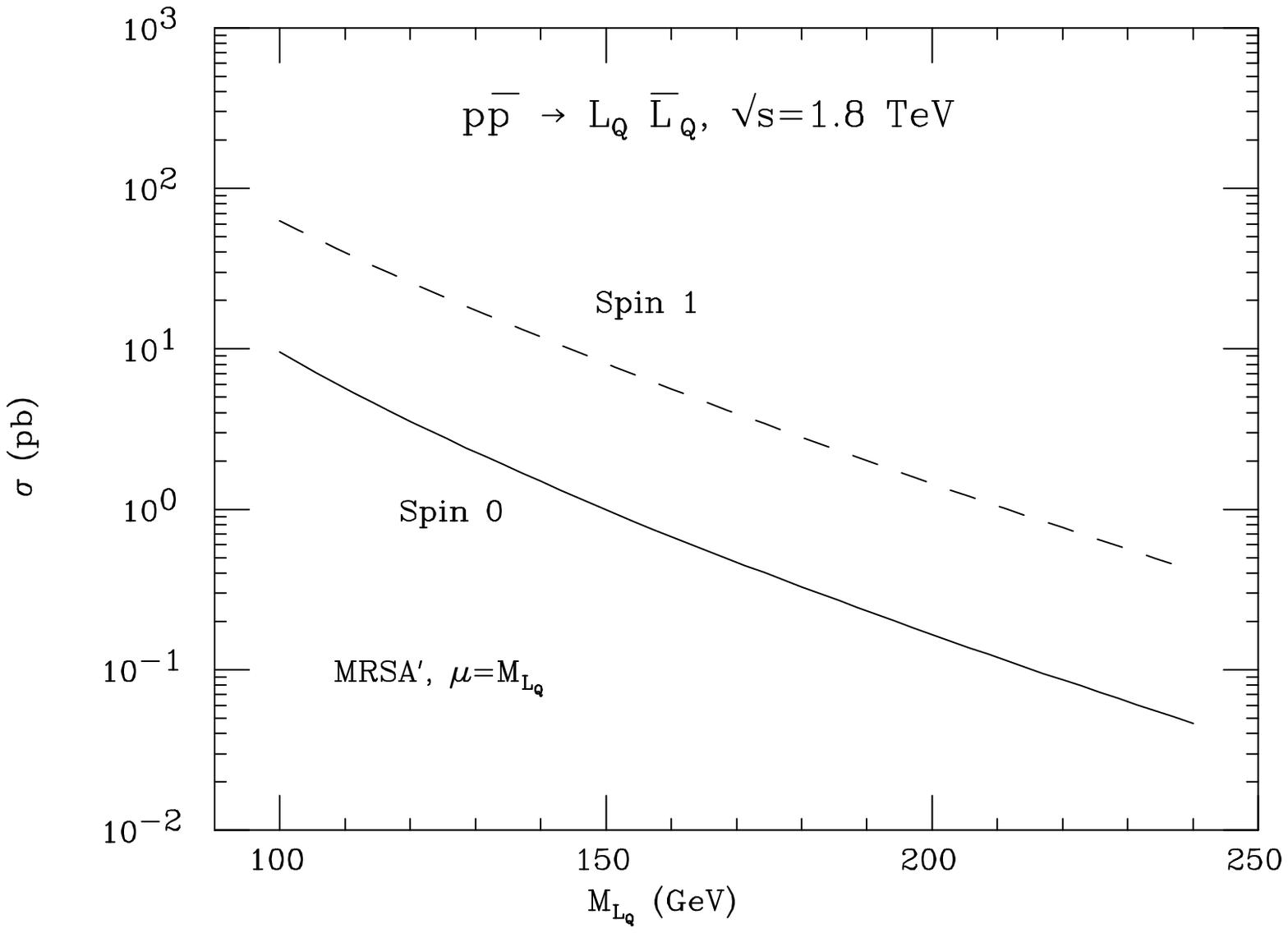,width=0.85\textwidth,clip=}}
\ccaption{}{ \label{fig:tevlq}    
Scalar and vector leptoquark production cross sections at the Tevatron. In the
case of the vector leptoquark, as discussed in the text, we have only
included the light $q\bar q$ annihilation processes.}
\end{figure}

It is important to explore the possible implications of the existence of
200~GeV leptoquarks at the Tevatron, since there leptoquarks are produced 
via model-independent QCD processes with potentially large rates. The
production cross section
for scalar, colour-triplet particles at the Tevatron is given as a function of
the mass in Fig.~\ref{fig:tevlq}, using the
MRSA$^\prime$~\cite{Martin95} 
parton distribution function set, and a renormalization/factorization
scale $\mu=m$.
The lowest-order production cross section for a pair of 200~GeV 
leptoquarks is
0.18~pb~\footnote{For our choice of renormalization scale, 
we also expect a multiplicative
$K$ factor of the order of 1.10 to 1.15 due to higher-order QCD
corrections. These have been evaluated in the case of supersymmetric scalar
quarks in~\cite{Spira96}. The case of leptoquarks can be recovered by
assuming a very large gluino mass. In this case, diagrams due to
four-squark operators which are not present in the leptoquark case
only appear in the gluon-fusion production channel,
which is significantly suppressed. }. The total integrated luminosity  
collected by the CDF and D0 experiments is of the order of 200~pb$^{-1}$, 
 which would yield approximately 36 events if the hypothetical
leptoquarks had a 100\% branching ratio into
electrons. The detection efficiencies for such a signal quoted by the CDF
and D0 experiments are each of the order of 20\%, resulting in about 7
detected
events in total. Although the expected 3.5 events per experiment are not
sufficient to exclude a leptoquark of 200~GeV in either CDF or D0
individually, the best current limit being 194 GeV by D0 as mentioned in 
the Introduction, it is likely that exclusion of a 200~GeV scalar
leptoquark with 100\% branching ratio into electrons
at more than the 95\% CL would result from 
the absence of a signal in a combined
analysis of the data collected by CDF and D0. For this reason, in the
following section we disfavour models with decay modes dominated by
electrons~\footnote{ {\em A priori}, generic leptoquarks could avoid this
problem by having two different Yukawa couplings, allowing $\nu q$ decays
as mentioned in the Introduction, or by coupling to $\mu q$ or  $\tau q$.
However, the $\mu q$ option is strongly constrained by limits on
flavour-changing
interactions, such as $\mu - e$ conversion on nuclei~\cite{Buch86,lqlimits}.}
although this possibility is not yet rigorously excluded.

The case of a 200~GeV vector leptoquark is most likely totally ruled out by the
Tevatron data, since the production rate can be as much as a factor
of 10 larger than that of                 
scalar leptoquarks, as shown in Fig.~\ref{fig:tevlq}. We emphasize that
in calculating this curve we have only included the light quark
annihilation
processes, and assumed minimal couplings to the gluons~\cite{Bluemlein93}. 
In the absence of a definite model of vector
leptoquarks, their coupling to gluons is not uniquely defined, and in general
leads to bad high-energy behaviour and unphysically large cross sections. This
problem has been studied in detail in refs.~\cite{Bluemlein96,Bluemlein97}, 
where $J=1$ leptoquark-pair production was considered for a general class of
anomalous couplings. In ref.~\cite{Bluemlein96,Bluemlein97} it was shown that,
even allowing for anomalous couplings, and selecting them so as to minimize the
production cross section via destructive interference, the total rate would
still be a factor of two larger than that for scalar leptoquarks. In the
absence of a signal, such a large rate would not be consistent with a combined      
CDF+D0 analysis if ${\cal B}(e^+ q) =1$.  If one discards the possibility of
such a fine-tuning in the anomalous couplings, values of ${\cal B}(e^+ q)$
significantly below 1 could be excluded.
                                                                      
\section{R-Parity Violation}
We find it attractive to embed a hypothetical leptoquark
in a well-motivated theoretical framework capable
of constraining its properties and providing other experimental
signatures, as is provided by the minimal supersymmetric extension
of the Standard Model~\cite{MSSM} with violation of $R$
parity~\cite{RPth}. The corresponding superpotential can be
written in the form
\beq
W_R \equiv \mu_i H L_i + \lambda_{ijk} L_i L_j E^c_k
+ \lambda'_{ijk} L_i Q_j D^c_k + \lambda ''_{ijk} U^c_i D^c_j D^c_k ~,
\label{WR}
\eeq
where $H, L_i, E^c_j, Q_k, (U,D)^c_l$ denote superfields for the
$Y= 1/2$ Higgs doublet, left-handed lepton doublets, 
lepton singlets, left-handed quark doublets and quark
singlets, respectively. The indices $i,j,k$ label the three generations
of quarks and leptons. Henceforth, we work in a basis for the $L_i$
and the $Y = - 1/2$ Higgs doublet $\bar H$ in which $\mu_i = 0$,
and the only surviving bilinear term is the Higgs mixing $\mu H {\bar H}$. 
Furthermore, we assume the absence of the $\lambda''$ couplings, so as to
avoid rapid baryon decay, and the $\lambda$ couplings play no r\^ole
in our analysis. 

\subsection{Production mechanisms}
The squark production mechanisms permitted by the $\lambda'$
couplings in (\ref{WR}) include $e^+ d$ collisions to form
${\tilde u}_L, {\tilde c}_L$ or $\tilde t_L$, which 
involve valence $d$ quarks, and various collisions of the
types $e^+ {d_i}$ ($i=2,3$) or $e^+ {\bar u_i}$ ($i=1,2,3$) which involve
sea quarks. 
The required magnitude of the coupling 
$\lambda'$ is fixed by the
observed product of the cross section $\sigma$ and the squark branching ratio 
$\cal B$ for the $R$-parity violating mode ${\tilde q}\to e^+ q'$. From
the results of the previous section, summarized in Table 2, we infer
that the valence production mechanism requires $\lambda'_{1j1}$ ($j=1,2,3$) 
to be about $0.04/\sqrt{\cal B}$, while any of the sea production
mechanisms require $\lambda'_{1jk}>0.3/\sqrt{\cal B}$ ($j,k=1,2,3$). The
latter are only marginally compatible with LEP~2 limits~\cite{Opal97},
and with previous H1 limits~\cite{H1limit} in the cases $j$ or $k = 1$.

The required values of the $\lambda'_{1jk}$ are to be compared with the
upper limits available from
various other laboratory experiments~\footnote{We will not consider here
very
stringent limits on $R$-parity violating interactions coming from
cosmological considerations of the baryogenesis energy
scale~\cite{coss}, since there are ways to avoid these in principle,
such as baryogenesis at the electroweak scale~\cite{EWBG}.}. It has been
inferred
from upper limits on neutrinoless $\beta \beta$ decay that \cite{Hirsch}
\beq
\vert \lambda'_{111} \vert < 7 \times 10^{-3} \left( {m_{\tilde q} \over 200 
~\hbox{GeV}}\right)^2      
\left( {m_{\tilde g} \over 1 ~\hbox{TeV}}\right)^{1 \over 2}~.
\label{betabeta}
\eeq
where $m_{\tilde q}$ is the mass of the lighter of
${\tilde u}_L$ and ${\tilde d}_R$, and $m_{\tilde g}$ is the gluino mass.
This limit excludes any production mechanism involving only first-generation
particles, and in particular the valence parton process $e^+d\to 
{\tilde u}_{L}$.

For charm squark production $e^+d\to 
{\tilde c}_{L}$, the most important constraint on the relevant coupling
constant $\lambda'_{121}$ comes from limits
on flavour-changing neutral current processes.
The simultaneous presence of several $\lambda'$ couplings with 
different flavour indices leads in general to dangerous tree-level
flavour violations. Usually one makes the most conservative assumption
that only a single $\lambda'$ coupling with specific flavour indices
is non-negligible. However, because of the mismatch in flavour space
between the up- and down-type left-handed quarks, this hypothesis
cannot be simultaneously satisfied both in the up and down sectors.
We will work in a basis where, in terms of the lepton ($N_i,E_i$) and
quark ($U_i,D_i$) mass eigenstates, the $\lambda'$ interaction term in
the superpotential is written as 
\beq
\lambda'_{ijk}(N_iV_{jl}D_l-E_iU_j)D^c_k~,
\eeq                   
where $V_{ij}$ is the usual Cabibbo-Kobayashi-Maskawa matrix. We will also
implicitly assume that the only sources of flavour violations are
described by $V$ and by the $R$-parity violating interactions.
Because of the non-trivial mixing in the down sector, the $\lambda_{ijk}$
couplings are bounded by
${\cal B}(K^+\to \pi^+ \nu \bar \nu )<2.4\times 10^{-9}$~\cite{BNL} to be
\cite{Agashe}
\beq
\vert \lambda'_{1jk} \vert 
< 2 \times 10^{-2} \left({m_{{\tilde d}_{k_R}} \over 200
~\hbox{GeV}}\right) \; \; \; \hbox{for} \; j= 1,2, \, k=1,2,3.
\label{dalign}
\eeq
Therefore a ${\tilde c}_L$ interpretation of the HERA data, which
implies $\lambda'_{121}\sim 0.04/\sqrt{\cal B}$, is possible if
\beq
m_{{\tilde d}_R}>\frac{400~{\hbox{GeV}}}{\sqrt{\cal B}} ~.
\label{b400}
\eeq
However, this bound on $m_{{\tilde d}_R}$ can be partially relaxed if
the mixing in the down sector is somewhat suppressed by the simultaneous
presence of various non-vanishing coupling constants $\lambda'_{ijk}$.
For instance, allowing for several $\lambda'_{1j1}$ with different indices
$j$, the bound in eq.~(\ref{dalign}) becomes
\beq
\sqrt{ \left| \sum_j \lambda'_{1j1} \frac{V_{j1}}{V_{21}} \right|
\left| \sum_l \lambda'_{1l1} \frac{V_{l2}}{V_{22}} \right|} <
2 \times 10^{-2} \left({m_{{\tilde d}_{R}} \over 200
~\hbox{GeV}}\right) ~.
\eeq
If $\lambda'_{111}$ saturates the bound in eq.~(\ref{betabeta}), then
$\lambda'_{121}$ can be as large as $4\times 10^{-2}$, even with
$m_{{\tilde d}_R}=200$ GeV. We want to stress that such a cancellation is
not necessarily accidental, but could arise as a consequence of a
particular
alignment of the $R$-violating interactions in flavour space. The simultaneous
presence of different couplings $\lambda'_{1j1}$ entails, in our basis,
some flavour violation in the up sector. We expect therefore new effects
in $D^0$--${\bar D}^0$ mixing and in the decay modes $D^0\to e^+ e^-$,
$D^+\to \pi^+ e^+ e^-$. These processes at present do not set constraints
on $\lambda'_{1j1}$ more stringent than the one considered
above~\cite{lqlimits}. We also
remark that the bound in eq.~(\ref{b400}) can also be
further relaxed by analogous cancellations among various
$\lambda'_{12k}$ couplings with different indices $k$.

In any case, the ${\tilde c}_L$ interpretation of the HERA data seems to
suggest that
${\cal B}(K^+\to \pi^+ \nu \bar \nu )$ is very close
to its experimental bound, a prediction which can be tested in the near
future in the ongoing Brookhaven experiment~\cite{BNL}.
Notice however that, while $m_{{\tilde d}_{R}}$ determines
the effective interaction responsible for $K^+ \to 
\pi^+ \nu \bar \nu$,
the HERA process is sensitive to $m_{{\tilde c}_L}$. 
In the absence of
a complete theory describing all supersymmetric particle masses, 
$m_{{\tilde d}_{R}}$ and $m_{{\tilde c}_L}$ 
are not necessarily related, and this prevents us from a
definite prediction for ${\cal B}(K^+\to \pi^+ \nu \bar \nu )$. 

Notice that one mass relation can be obtained
in the case of the ${\tilde c}_L$ interpretation of
the HERA data, with
the help of weak-SU(2) symmetry. Indeed, assuming
no significant left-right squark mixing, 
 we can predict 
$m_{{\tilde s}_L} = \sqrt{m^2_{{\tilde c}_L} - \hbox{cos} 2 \beta M_W^2
} \sim 200$ to 220 GeV. 
The squark ${\tilde s}_L$ cannot be produced at HERA from valence parton
processes, but could be observed at the Tevatron.

The last possibility of a
valence production mechanism is
$e^+ d \rightarrow {\tilde t}_L$ via $\lambda'_{131}$.
Apart from the H1 and OPAL limits mentioned earlier, 
this coupling constant
is constrained by
experiments on parity violation in atomic physics, which
imply \cite{noi}                   
\beq
\vert \lambda'_{131} \vert < 4\times 10^{-1} 
\left({m_{{\tilde t}_L} \over 200 ~\hbox{GeV}}\right)
~.
\label{pvatoms}
\eeq
This certainly allows a sufficient production rate,
even if $\cal{B}$ is significantly less
than 1.
In the absence of left-right stop mixing, one expects
$m_{{\tilde b}_L} = \sqrt{m^2_{{\tilde t}_L} - \hbox{cos} 2 \beta M_W^2
- m_t^2 + m_b^2} \sim 100$ to 130 GeV, in which case there
would be an excessive contribution to the electroweak $\rho$ parameter:
$\Delta \rho \sim 2$ to $4 \times 10^{-3}$. 
At present the experimental value of $\Delta \rho$ from LEP/SLD data
plus $m_W/m_Z$ measurements is $\Delta \rho_{exp} =(4.7\pm 1.3)\times
10^{-3}$ \cite{alt}. The Standard Model value for $m_t=175$ GeV is
$\Delta \rho_{SM}=5.7 \times 10^{-3}$ and $4.9 \times 10^{-3}$ 
for $m_H=100$ and $300$ GeV,
respectively. Thus, there is little space for a new positive contribution
to $\Delta \rho$.
This suggests the
necessity of significant left-right stop mixing, which is not unnatural
and could also accommodate the lightness of the stop with respect
to the other squarks. In this case, ${\tilde b}_L$ could
be heavier and $\Delta \rho$ reduced. This also entails that the
value of $\lambda'_{131}$ inferred naively from the HERA data is smaller 
than the actual value
by a factor of $\cos \phi_t$, where $\phi_t$ is the stop mixing angle.

Most sea production processes are excluded by a combination
of different
experimental
constraints.
Some of these have been discussed above; others come from
contributions to the electron neutrino mass \cite{neutr}
\beq
\vert \lambda'_{133} \vert < 5 \times 10^{-3} \left({m_{\tilde q} \over 200~ 
\hbox{GeV}}\right)^{1
\over 2}\; \quad , \quad \vert \lambda'_{122} \vert 
< 1 \times 10^{-1} \left({m_{\tilde q} 
\over 200~\hbox{GeV}}\right)^{1 \over 2}~,
\label{numass}
\eeq
which exclude sea-quark production mechanisms involving only third- (or only
second-) generation 
particles. In eq.~(\ref{numass}) we have assumed a common 
supersymmetry-breaking mass $m_{\tilde q}$ in the $2\times 2$ mass matrix
for the ${\tilde d}_L$--${\tilde d}_R$ system, and taken $m_s(m_{\tilde q})
=100$ MeV for the running strange quark mass. Finally,
${\tilde u}_L$ production off sea quarks of
the second or third generation is constrained by
limits on charged-current
universality, which impose \cite{noi}
\beq
\vert \lambda'_{11k} \vert 
< 6 \times 10^{-2} \left({m_{{\tilde d}_{k_R}} \over 200
~\hbox{GeV}}\right) \; \; \; \hbox{for} \; k=1,2,3.
\label{ccuniv}
\eeq
Also, if the observed anomaly is due to
the sea process $e^+ \bar u \to \bar{\tilde{d}}_{k_R}$, then an effect
more than 50 times larger should have shown up in 
$e^-  u \to {\tilde{d}}_{k_R}$, while no anomaly has been observed
in about 1 pb$^{-1}$ of data collected in $e^- p$ collisions \cite{deep}.

The only remaining possibility for production on sea partons is
$e^+s\to {\tilde t}_L$ via the $\lambda'_{132}$ interaction, which is
constrained, but not quite excluded, by the OPAL analysis~\cite{Opal97}.
Limits on
anomalous top quark decay modes 
also set weak bounds on this coupling \cite{Agashe},
but these disappear as soon as the selectron mass is not much
smaller
than $m_t$. This interaction can also give new contributions to the $b\to 
s \gamma$ decay rate, but these effects can be suppressed by an
approximate alignment in the down sector. Therefore $\lambda'_{132}$ could
be as large as 0.3 and the process $e^+s\to {\tilde t}_L$ be at the
origin of the HERA signal. In this case, as we will discuss in the next
section, 
the scattering process $e^+e^-\to {\bar s}s$ has an anomalous
contribution due to $t$-channel stop exchange which can be easily identified
at future LEP~2 runs.

\subsection{Decay patterns}
Next we address the issue of the squark decay modes. In the
case of ${\tilde c}_L$, the
most important possible decay modes are 
the $R$-conserving ${\tilde c}_L\rightarrow c \chi^0_i$ ($i=1,..,4$) and
${\tilde c}_L\rightarrow s \chi^+_j$ ($j=1,2$), and 
the $R$-violating ${\tilde c}_L\rightarrow d e^+$, where
$\chi^0_i, \chi^+_j$ denote neutralinos and charginos, respectively.
If $R$-parity violating couplings other than $\lambda'_{121}$ were
present, further decay modes could be allowed, although this possibility
is severely constrained by limits on
flavour and lepton conservation. 
The decay rate for the $R$-parity violating mode is ~\cite{Hewett,Dreiner,Kon} 
\beq
\Gamma ({\tilde c}_L \to e^+ d ) = {1 \over 16 \pi} (\lambda'_{121})^2 
m_{{\tilde c}_L}
\label{derate}
\eeq
and the coupling $\lambda'_{121}$ is fixed by 
our production assumption. It has often
been found that the $R$-conserving modes dominate, but this
is not
necessarily the case. They could be either suppressed by phase space
or, in the ${\tilde c}_L\to c \chi^0_i$ case, 
by (partial) cancellations in the neutralino
couplings. The $s \chi^+_j$ decay mode can only be suppressed by
phase space, so we assume that $m_{\chi^+_j} > 200$ GeV.
Neglecting ${\tilde c}_{L,R}$ mixing, 
the decay rate for the neutralino
mode is given by ~\cite{Dreiner}
\beq
\Gamma ({\tilde c}_L \to c \chi^0_i ) = 
\frac{g^2}{32 \pi} (A_i^2+B_i^2)~ m_{\tilde{c}_L} \left
(1-\frac{m^2_{\chi^0_i}}{m_{\tilde{c}_L}^2} \right )^2 ~,
\label{chirate}
\eeq
where
\beq
A_i = \frac{ m_c N_{i4}}{ M_W \sin\beta}, \; \; \;
B_i = 
N_{i2} + \frac{1}{3} \tan \theta_{W} N_{i1} ~.
\label{cancel}
\eeq
In eq.~(\ref{cancel}) 
the $N_{ij}$ are the elements of the unitary matrix
that diagonalises the neutralino mass matrix
in the SU(2) - U(1) gaugino basis~\cite{MSSM}, and $\tan \beta$ is the
ratio of Higgs vacuum expectation values. The quark-mass-suppressed
$A_i^2$
term in eq.~(\ref{chirate}) may be neglected, and we notice that the
$B^2_i$ term in eq.~(\ref{chirate}) is reduced
either if the lightest neutralino is an approximate higgsino
($N_{11}\sim N_{12} \sim 0$) or
if there is a cancellation
\beq 
N_{12} \sim - {1 \over 3} \hbox{tan} \theta_W N_{11} ~.
\label{condition}
\eeq
Remarkably enough, we find that the cancellation (\ref{condition})
does occur in an acceptable domain of supersymmetric parameter space,
given analytically by
\beq
m_{\chi^0_1}=\frac{4 \sin^2 \theta_W}{3-4\sin^2\theta_W}M_2
\label{Gian}
\eeq
\beq
\mu =\sin 2 \beta X \pm \sqrt{\sin^2 2 \beta X^2 +m_{\chi^0_1}(m_{\chi^0_1}
+2X)}
\eeq
\beq
X\equiv \frac{2(1-\sin^2\theta_W)(3-4\sin^2\theta_W)}{(3-8\sin^2\theta_W)}
\frac{M_Z^2}{M_2}~.
\eeq
Here $M_2$ is the SU(2) gaugino mass, while the U(1) gaugino mass is
determined by the unification relation $M_1=(5/3)\tan^2\theta_WM_2$.
Notice that such a cancellation is possible for ${\tilde c}_L\to c \chi^0_1$,
but impossible,
for instance, in the
analogous ${\tilde d}_R$ decay, whose rate is still given by 
eq.~(\ref{chirate}), with $B_i =-\tan \theta_W N_{i1}$.

\begin{figure}
\centerline{\epsfig{figure=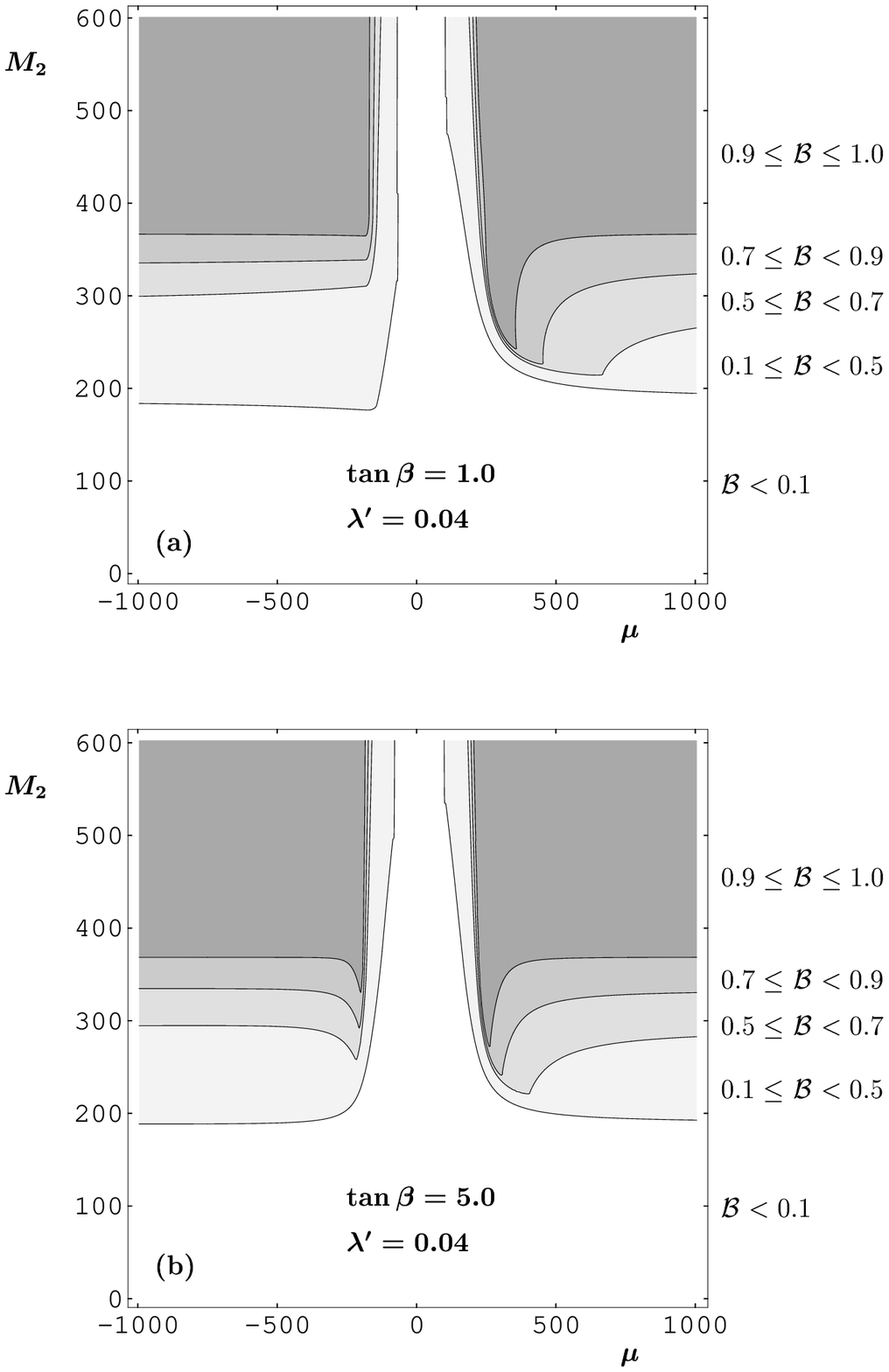,width=0.7\textwidth,clip=}}
\ccaption{}{ \label{fig:charmbr}    
Contours of ${\cal B}(e d)$ for the $R$-violating decay of $\tilde
c_L$, in
the $\mu-M_2$ plane. Here $\lambda^{\prime}$ has been fixed to 0.04 and
$\tan\beta=1$ (a) and $5$ (b), respectively. The LEP~2 bound of 85 GeV for
the chargino
has also been implemented.}                 
\end{figure}                                                       

\begin{figure}
\centerline{\epsfig{figure=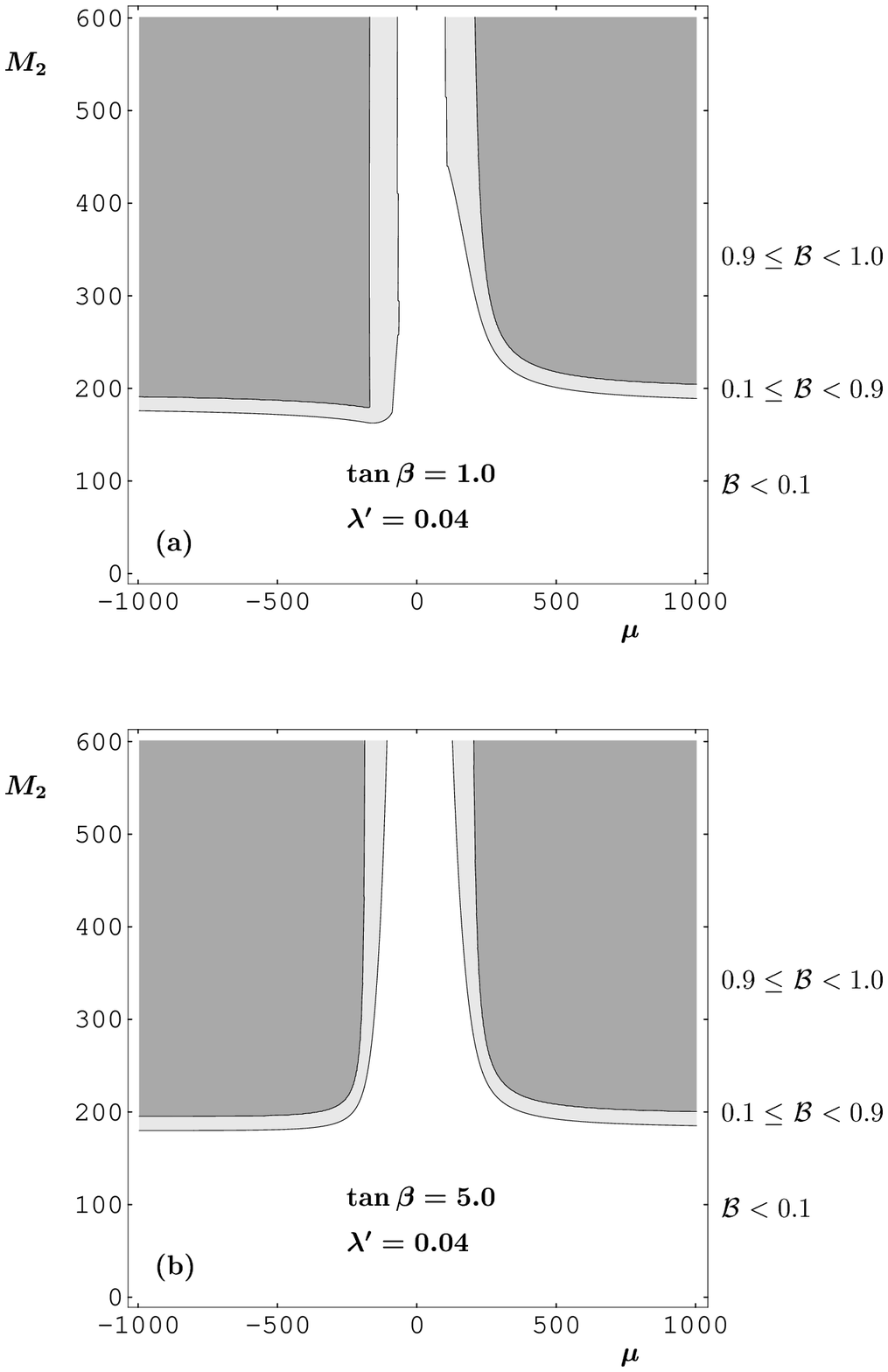,width=0.7\textwidth,clip=}}
\ccaption{}{ \label{fig:topbr}    
Contours of ${\cal B}(e d)$ for the $R$-violating decay of $\tilde t_L$,
in
the $\mu-M_2$ plane. Here $\lambda^{\prime}$ has been fixed to 0.04 and
$\tan\beta=1$ (a) and $5$ (b), respectively. We have assumed a vanishing
stop left-right mixing. The LEP~2 bound of 85 GeV for
the chargino
has also been implemented.}                 
\end{figure}                                                       
     
The results of numerical studies, see Fig.~\ref{fig:charmbr}a, explicitly
show
the three regions where the ${\tilde c}_L$ $R$-parity violating
decay modes become important. The first region occurs for $M_2$ and
$\mu$ large enough to suppress kinematically all two-body $R$-parity
conserving modes ($m_{\chi^0},m_{\chi^\pm}>m_{{\tilde c}_L}$).
In this case, ${\cal B}=1$ and $\tilde{c}_L$ should lie at the edge of the
parameter region excluded by D0.       
The second region is the thin slice of parameter space where $\chi^0$
is an approximate higgsino ($\mu << M_2$). In this case, the couplings of
$\tilde{c}_L$ to the light chargino and neutralinos are suppressed, and the
$R$-parity violating mode can compete with the $R$-parity conserving ones.
In the third region, the 
decay mode ${\tilde c}_L \to c \chi^0$ is suppressed by the approximate 
cancellation of eq.~(\ref{condition}). This
cancellation is especially
marked for tan$\beta \sim 1$, where ${\cal B} > 0.5$ over
a large domain of $\mu$.
The extent
of the cancellation region is reduced as tan$\beta$ is increased,
as can be seen from Fig.~\ref{fig:charmbr}b, 
which is for tan$\beta = 5$, since the
region where eq.~(\ref{condition}) is approximately satisfied becomes
narrower as tan$\beta$ increases. We conclude from Fig.~\ref{fig:charmbr}
that the detection of $e^+ q$ final states 
by H1 and ZEUS data should not be a surprise.

A small value of ${\cal B}$ could 
in principle be accommodated
by increasing the magnitude of the $\lambda'_{121}$ coupling by the
corresponding factor of $1 / \sqrt{\cal B}$, though  
the scope for this is severely
limited by the bounds described above.
On the other hand, if $\cal B \sim$ 1, the squark
should be at the verge of being discovered at the Tevatron or
possibly ruled out by CDF and D0 data, as discussed in the previous section.

In the case of ${\tilde t}_L$,            
it is interesting to notice that the
neutralino decay mode ${\tilde t}_L \to t \chi^0_i$ is kinematically closed 
in a natural way. In order to obtain a large value of $\cal B$, it is 
sufficient to require that all charginos are heavy enough to forbid the
decay ${\tilde t}_L \to b \chi^+_j$ (see figs.~\ref{fig:topbr}a and 
\ref{fig:topbr}b). In view of the
Tevatron limits discussed in
the previous section, that may pose a problem if ${\cal B}(e q)$ is very
close to 1, we have analyzed other possible decay modes.
Under the gaugino unification assumption with $m_{\chi^+_j}>m_{{\tilde
t}_L}$,
the three-body decay ${\tilde t}_L \to b W^+\chi^0_{i}$ (mediated by a
virtual $\chi^+_j,t$ or ${\tilde b}_L$) has a negligible rate. Even if
a slepton were much lighter than the stop, the decay modes ${\tilde t}_L
\to b\ell^+ {\tilde \nu}$, $b \nu {\tilde \ell}^+$ could not compete with
the $R$-parity violating decay, for $\lambda'_{131}=4\times 10^{-2}$.
It is usually assumed that the flavour-violating decay ${\tilde t}_L \to
c \chi^0_i$ is rather suppressed in supersymmetric models. However,
in theories with $R$-parity violation, the whole issue of flavour
conservation is undermined, and we cannot exclude new unexpected
effects, which could lead to large rates for ${\tilde t}_L \to
c \chi^0_i$ and values of $\cal B$ considerably smaller than 1.

If the stop is produced by the sea-parton collision $e^+ s\to {\tilde t}_L$,
then the $R$-parity violating decay is fast enough to compete
with the chargino mode. The $R$-violating branching ratio $\cal{B}$
depends only on the chargino masses $m_{\chi^+_j}$ and their gaugino
compositions $|V_{j1}|$. We find ${\cal B} =0.5$ either for a pure gaugino-like
chargino ($|V_{11}|=1$) of 150 GeV, or for a mixed chargino ($|V_{11}|=
1/\sqrt{2}$) of 120 GeV. Therefore, in this case, it is easier to
escape the Tevatron limits, although a small value of $\cal{B}$ requires
a large value of $\lambda'_{132}$ and a large effect in $e^+e^-\to
{\bar s}s$ at LEP~2.

\begin{figure}
\centerline{\epsfig{figure=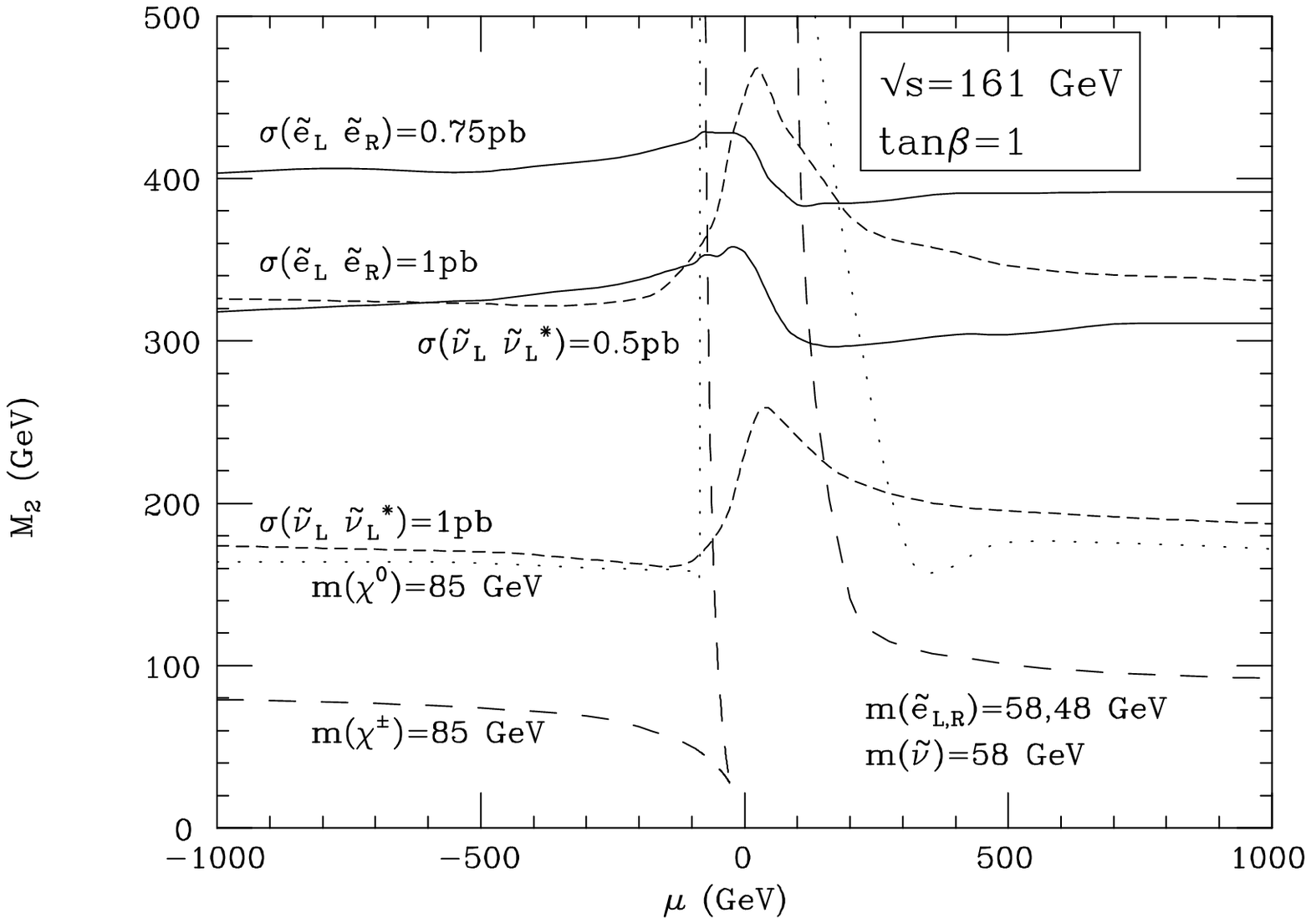,width=0.85\textwidth,clip=}}
\ccaption{}{ \label{fig:aleph1}    
Crosss section contours for 
${\tilde e}_L {\tilde e}_R$ and ${\bar {\tilde \nu}} {\tilde \nu}$
production at LEP~2 ($\sqrt{s}=161$ GeV and $\tan\beta=1$). 
The calculations were performed using the programs {\tt SUSYXS} documented in
ref.~\cite{Mangano96}, and include the effect of ISR.
The regions                                          
corresponding to lightest chargino and neutralino masses heavier than
85~GeV lie above the dashed and dotted lines, respectively.}
\end{figure}

It is natural to ask whether the $R$-violating scenario for the HERA
events discussed above is compatible with the $R$-violating scenario
proposed elsewhere~\cite{CGLW} to interpret the four-jet excess seen by
ALEPH \cite{ALEPH} at
LEP~2~\footnote{We are aware that the ALEPH signal~\cite{ALEPH} has not
been confirmed 
by the other LEP collaborations~\cite{Schlatter}, but reserve judgement on
the final fate
of the four-jet excess.}. The suggestion contained in ref.~\cite{CGLW} 
was that ALEPH
had observed the $R$-conserving production process $e^+ e^- \rightarrow
{\tilde e}_L {\tilde e}_R$, mediated by an approximate 
U(1) gaugino with mass
$M_1$ between 100 and 120 GeV. The subsequent $R$-violating decays of each
${\tilde e}_{L,R}$ into ${\bar q} q$ produce a pair of hadronic jets
via an interaction of the $\lambda'$ type. This could have the
$121$ flavour structure advocated above for ${\tilde c}_L$ production at 
HERA, in which case we predict the
presence of a charm quark and an antiquark in the two
dijets of the ALEPH final states. 
A strong constraint on this scenario comes from an adequate suppression
of sneutrino pair production, which requires $\tan \beta \sim 1$ and $M_2$
larger than what predicted by gaugino unification. It is interesting to notice
that such a choice of parameters can also lead to the approximate cancellation
described in eq.~(\ref{condition}), and therefore to a significant value for
$\cal{B}$. 
We display in
Fig.~\ref{fig:aleph1} contours of the cross sections for 
${\tilde e}_L {\tilde e}_R$ and ${\bar {\tilde \nu}} {\tilde \nu}$
production at LEP~2 for $\sqrt{s}=161$ GeV and $\tan\beta=1$.
We do not show the corresponding figure for larger values of $\tan \beta$
since, for $m_{{\tilde e}_L}=58$ GeV, values of $\tan \beta >1.23$ are
excluded by the requirement $m_{\tilde \nu}>M_Z/2$. In Fig.~\ref{fig:aleph1} 
we have assumed gaugino mass unification, in order to allow the
comparison with the results shown in Fig.~\ref{fig:charmbr}. However,
as mentioned above, a better agreement with the ALEPH data is actually
obtained when $M_2$ is larger than $(3/5)\tan^{-2}\theta_W M_1$.
We conclude that the $R$-violating interpretation of the HERA data in
terms of ${\tilde c}_L$ production
advocated above is not incompatible
with that proposed previously~\cite{CGLW} for the ALEPH four-jet events.

\section{Tests to Discriminate between Models}
In this final section we review some key experimental tests that
may help to distinguish between different novel physics
interpretations of the HERA large-$Q^2$ events. 

The first comment follows from Fig.~\ref{fig:complim}: the $Q^2$
distribution expected from effective contact interactions and resonance
interpretations are different and could be distinguished clearly with a
modest increase in statistics. The present data seem to favour a resonance
interpretation, but it would be premature to draw firm conclusions at this
stage. As for the $x$ distributions, more statistics are again
required to establish consistency with a resonant peak smeared out
by ISR (see the Appendix), gluon radiation, hadronization and detector
effects.

``Charged current" events due to $\nu q$ decays would be expected in some
leptoquark and $R$-violating squark scenarios at rates similar to those
for the ``neutral current" events seen. However, this is not the case,
in particular, for the valence production $e^+ d \rightarrow {\tilde
c}_L/{\tilde t}$ scenario favoured above.
The H1 collaboration has reported~\cite{H1} four ``charged current" events
in a kinematic region where less than two are expected according to the
Standard Model. The scenario we favour would be excluded if this
signal built up into a significant signal with the advent of more data.

The recorded luminosity in $e^- p$ collisions is rather limited,
although it has already provided some constraints on scenarios in
which a leptoquark ($R$-violating squark) is produced via $e^+$ 
collisions with a sea quark, as commented in the previous section.
A much higher $e^- p$ integrated luminosity should become
available in the future. The cross section curves shown in
Fig.~\ref{fig:lqxsec} are also applicable to these collisions.
It is a key prediction of our preferred ${\tilde c}_L/{\tilde t}$ model
that there should
be no large cross section for the production of a resonance peak in
$e^- p$ collisions.

One of the options for future HERA running is for $e^+ D$
collisions~\cite{Arneodo96}. This is potentially interesting, since the HERA
signal is made from $e^+ d$ collisions according to our favoured
interpretation, the neutron
contains twice as many $d$ quarks as the proton, and this ratio
is further enhanced at large $x$. The luminosity for $e^+ d$
collisions in scattering off a neutron is the same as that for
$e^+ u$ collisions in a proton, and can be read off from Fig.~\ref{fig:lqxsec}.
Unfortunately, the effective $E_{CM}$ in $e^+ n$ scattering will
be less than the 300 GeV currently attained with protons.

We recall that, for the reasons discussed above, the ${\tilde c}_L
\rightarrow c (\chi^0 \rightarrow {\bar q} q \ell, \nu)$ decay
chain may have a branching ratio comparable to the $e^+ q$
final state that we hypothesize to have been observed, and
these should be observable at HERA. We note
that $\ell^{\pm}$ final states are equally likely in $\chi$
decays, and that dominance by $\ell = \mu, \tau$ cannot be
excluded. These final states would all have very clear signatures:
charged leptons which may well have different flavour and/or
charge from the incoming $e^+$, or missing energy
carried away by a neutrino, each accompanied by three hadronic
jets, at least one of which should contain a charmed particle.

\begin{figure}                                            
\centerline{\epsfig{figure=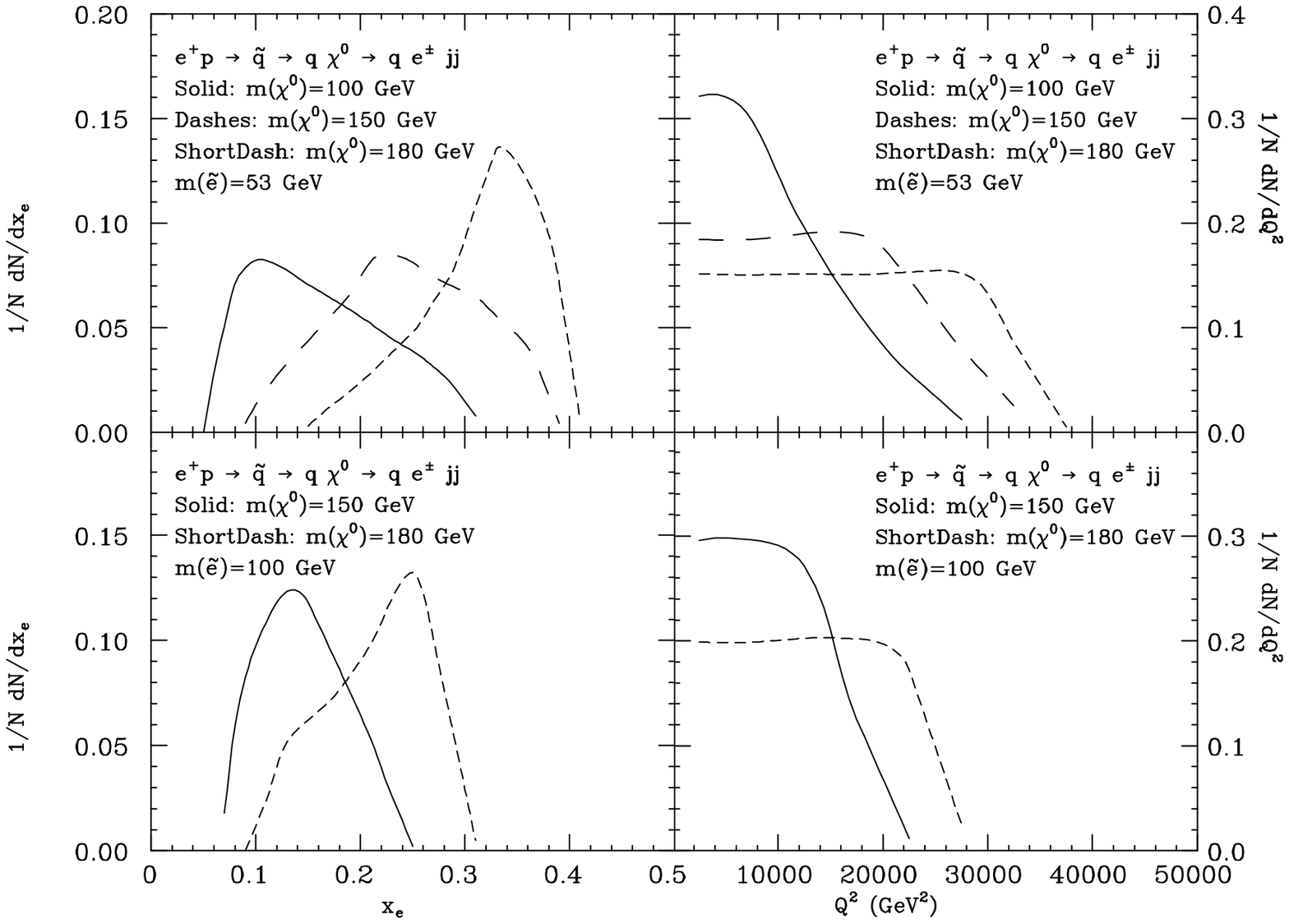,width=0.85\textwidth,clip=}}
\ccaption{}{ \label{fig:heradec}    
Kinematical distributions of wrong-sign electrons in the $R$-conserving
decays of squarks, followed by $R$-violating neutralino decay. Different
combinations of neutralino and slepton masses are
shown.}                                             
\end{figure}

In Figure~\ref{fig:heradec} we show the shapes of the $x_e$~\footnote{$x_e$ is
the Bjorken $x$ variable extracted using the electron method (see
refs.~\cite{H1,ZEUS}).} and $Q^2$
distributions for wrong-sign leptons due to $R$-conserving decays of
squarks at HERA. We have not implemented any selection cut
in preparing this figure, with
the exception of a $Q^2>5000$ GeV requirement for the events in the $x_e$
distribution. Small Standard Model
backgrounds to these final states are expected to come from semileptonic decays
of heavy quarks (charm and bottom). The separation of these backgrounds depends
significantly on the lepton isolation requirements, 
which are related to the detector characteristics and will not be studied
here. 
Since the distributions depend significantly on the masses of the states
produced in the decay chain, we consider two values of the slepton mass (53 and
100 GeV) and three values of the neutralino mass (100, 150 and 180 GeV). As
expected, the figures show that the most interesting signals arise in the case
of the largest neutralino masses and smaller slepton masses.

Our analysis has pointed up the urgency of a joint analysis of
the CDF and D0 dielectron data, to see whether 
the existence of a 200~GeV leptoquark (or
$R$-violating squark) can be probed with the available
Tevatron data, and, in the absence of a signal, down to what
${\cal B}(e q)$ it can be excluded. 
If the existence of such a leptoquark (squark) is still
a live issue at the time of the next Tevatron run, we expect that
the data taken then should be able to probe its existence down
to values of ${\cal B}(e q)$ that are below those of interest to
the present HERA data.

\begin{figure}
\centerline{\epsfig{figure=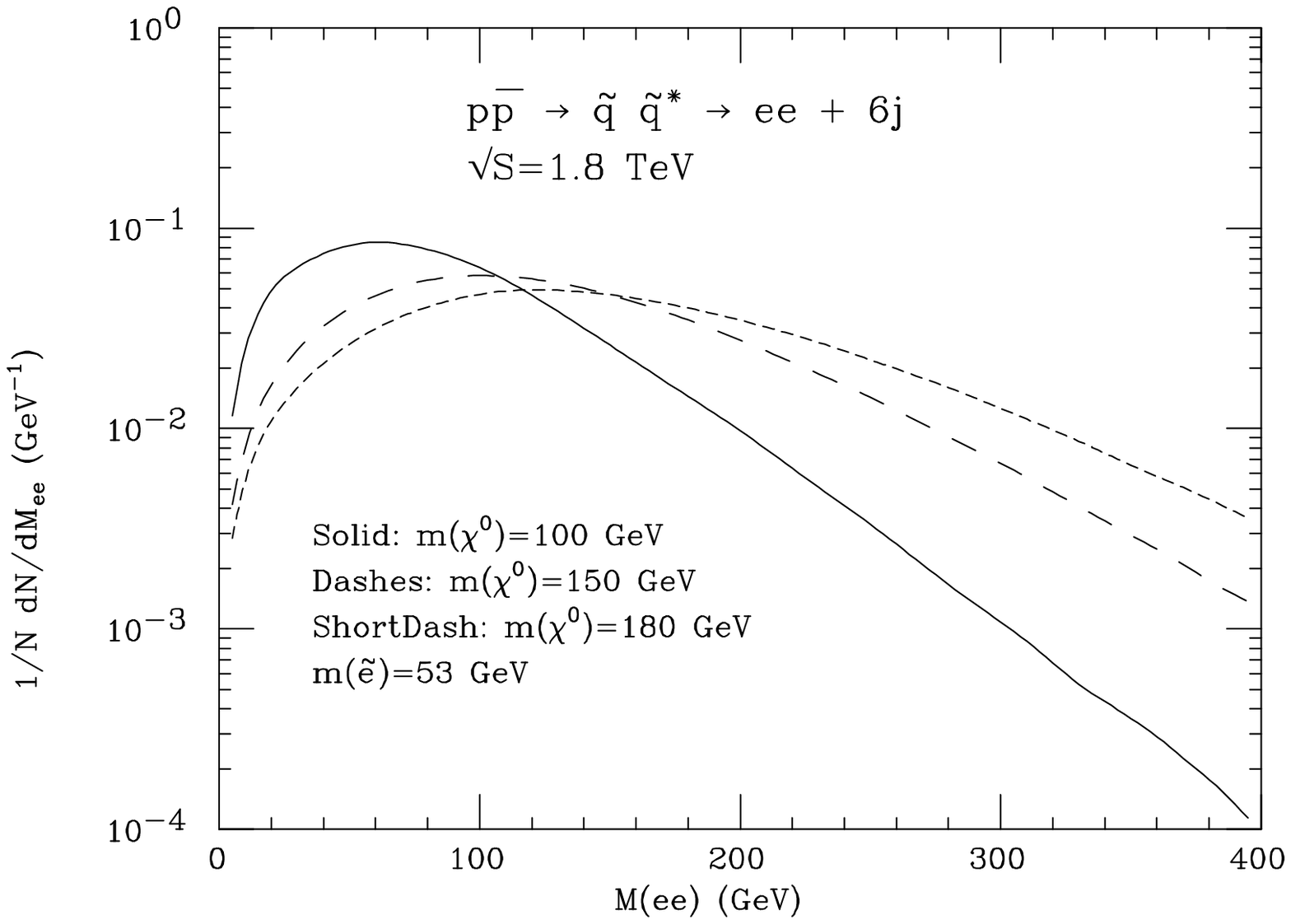,width=0.85\textwidth,clip=}}
\ccaption{}{ \label{fig:fnaldec}    
Invariant mass distributions of electron pairs from the $R$-conserving
decays of squark pairs produced at the Tevatron,
followed by $R$-violating neutralino decays. Different combinations
of neutralino masses are shown.}
\end{figure}

The cascade decays of the squark via an $R$-conserving
${\tilde c}_L \rightarrow \chi^0 c$ 
interaction could also have distinctive
signatures at the Tevatron. We show in
Fig.~\ref{fig:fnaldec} the invariant mass
distribution of lepton pairs, which is independent of the relative
charge of the leptons. The final states consist of two leptons (with equal
probability of having same or opposite charge), a large
number of jets (there are 6 energetic
quarks in the final state), and no missing
transverse energy~\footnote{If there are also $\chi \rightarrow \nu
{\tilde \nu}$
decays, more signatures would appear, such as $e^{\pm}$ + jets + missing
transverse energy.}.
Possible backgrounds come from the production and decay of
top quark pairs, when the net missing transverse energy 
carried away by neutrinos is small and additional jet are produced by radiative
processes.
In the case of like-charge leptons, the $t\bar{t}$
background requires one of the leptons to come from the semileptonic decay of
one of the $b$ quarks in the final state. Once again, the precise size of the
background will strongly depend on the lepton isolation requirements, and
will
not be estimated here.

In the case of stop production and decay via a virtual chargino, the signatures
become particularly interesting, because of the presence of $b$ jets in
the
final state. Transverse missing energy would also arise from the chargino
decay to neutrino and slepton. In the case of ${\tilde c}_L$ 
decays to $c\chi^0$ there will be charm
jets, which will still give rise to secondary vertices. Only higher statistics
will however allow this signal to be isolated and distinguished from the
secondary vertex distribution of $b$ decays.

It is natural to ask whether a squark weighing 200 GeV might have
some observable indirect effects at LEP 2, even though it could not be
produced directly. An $R$-violating interaction $\lambda'$ gives
a contribution to the cross section of a generic quark-antiquark
final state ${\bar f} f$ via the exchange of a squark $\tilde q_L$ or
$\tilde q_R$,
which is parametrised as
\begin{equation}
\sigma = \sigma_{SM} + \frac{ 3 \lambda'^4 I_1}{64 \pi s} +
\frac{ 3 \lambda'^2 \alpha_{em}I_2}{4 s} \left[
 e_e e_f + 
 a_L^e a_{L,R}^f             
\frac{s(s-M_Z^2)}{   
(s-M_Z^2)^2 + \Gamma_Z^2 M_Z^2}  \right ]
\label{dev}
\end{equation}

where
\begin{eqnarray}
I_1 & = & \frac{1+2x_{\tilde{q}}}{1+x_{\tilde{q}}} 
    -2x_{\tilde{q}}\ln\left( \frac{1+x_{\tilde{q}}}{x_{\tilde{q}}}\right)
\\  
I_2 & = & \frac{1}{2}-x_{\tilde{q}}+x_{\tilde{q}}^2\ln 
    \left( \frac{1+x_{\tilde{q}}}{x_{\tilde{q}}}\right)
\end{eqnarray}
with
$a_{L,R}^f = (T_3^f - e^f \sin^2\theta_W)/ (\sin\theta_W
\cos\theta_W)$ and 
$x_{\tilde{q}} \equiv \frac{m^2_{\tilde{q}}}{s}$. In the case of
$d \bar{d}$ production by ${\tilde c}_L$ exchange, we have
$a_R^d = -e_d \tan \theta_W$, whereas in the case of
${\bar c} c$ production by ${\tilde d}_R$ exchange we have
$a_L^c = (1/2 - 2/3\hbox{sin}^2 \theta_W)/(\hbox{sin}\theta_W
\hbox{cos} \theta_W)$.
The contribution in eq.(\ref{dev}) proportional to $I_1$
arises from the diagram with the R-parity violating     
vertices, while those proportional to $I_2$ are interference terms with
the Standard Model s-channel $\gamma$ and $Z$ exchange
respectively. For $\sqrt{s} = 192$ GeV, the correction to
the Standard Model cross section is $\approx 0.02$ pb, which we
suspect that is rather small to be observed.
However, there could be an observable signal in $e^+ e^- \rightarrow
{\bar s} s$ due to $\tilde t$ exchange in the sea production scenario
discussed above, which is already on the verge of exclusion by
OPAL~\cite{Opal97}.

Since the lightest neutralino $\chi^0$ decays rapidly via $R$-violating
interactions, the reaction $e^+ e^- \rightarrow \chi^0 \chi^0$ should be
observable at LEP~2 for $m_{\chi^0}$ below the kinematic limit, currently
about 85 GeV. We do not discuss here the production cross section, which
depends, e.g., on the selectron masses assumed. We have plotted in 
Fig.~\ref{fig:aleph1} 
the $m_{\chi^0} = 85$ GeV contours. Comparison with Figs.~\ref{fig:charmbr}
and \ref{fig:topbr}
indicates that our favoured $R$-violating HERA scenarios are not strongly
constrained by present LEP~2 data, though future data at $\sqrt{s} = 200$
GeV
might make some inroads on the parameter space.

The $R$-violating scenario squark mentioned in the previous section may
have observable consequences for other experiments that are not {\it a
priori} related. One example is $K \rightarrow \pi {\bar \nu} \nu$
decay~\cite{Agashe}.
We have seen that this imposes one of the most stringent constraints on 
the $\lambda'_{121}$ coupling that we invoke. A corollary is that there
may be an interesting contribution to this decay from beyond the Standard
Model, waiting to be discovered just below the present level of
experimental sensitivity.
The magnitude of any such signal depends on the pattern of flavour mixing
among $R$-violating couplings. In some variations, there may also be
contributions to nuclear $\beta\beta$ decay lurking just below the
present experimental sensitivity~\cite{Hirsch}.

These examples point to a general theoretical issue raised by the
possibility of $R$ violation. General $R$-violating couplings do
not respect the classic conditions for natural conservation of
flavour in neutral interactions~\cite{GWP}. Perhaps these are in any case
optional,                             
and one should be content with models which fall numerically below the
experimental upper limits on flavour-changing neutral interactions. On
the other hand, natural respect for them played an important
historical r\^ole in motivating the Standard Model. Therefore,
it is desirable to clarify whether there are any
interesting and plausible conditions under which these constraints
are naturally respected by $R$-violating interactions~\cite{Banks,Cham}.

This example shows that interesting and relevant theoretical, as
well as experimental, issues are into new light by the observation
of large-$Q^2$ events at HERA. It may well be that these turn out to
be a malign statistical fluctuation, rather than a harbinger of new
physics. However, we have shown in this paper that complementary
experiments may soon be able to cast light on possible
interpretations in terms of physics beyond the Standard Model. In
the mean time, the HERA large-$Q^2$ events have caused us to look
anew at supersymmetry with $R$ violation, in particular, and
provided us with new reason to question the conventional $R$-conserving
paradigm for supersymmetric phenomenology.
\\[0.5cm]
\noindent
{\bf Acknowledgements.} One of us (MLM) would like to thank A. Staiano for
several illuminating discussions on the details of the ZEUS experimental
analysis. We thank J. Bl\"umlein for pointing out to us the existence of
ref.~\cite{Bluemlein96} and A.~Nelson for a very useful remark.
The work of SL is funded by a Marie Curie Fellowship
(TMR-ERBFMBICT 950565).
\\[1cm]
\noindent {\large \bf Note Added}\\[0.3cm] \noindent
After the completion of this paper we received 
the articles in ref.~\cite{added1,added2}, which make the interesting
observation that experiments on
atomic parity violation
pose stringent limits on the strength of any
$A_eV_q = RR-LL+RL-LR$ $ \bar ee
\bar q q$ contact
term. The present experimental value on the coefficient $C_{1q}$ of the
$A_eV_q$ term in
the effective lagrangian, as given on pages 87-92 of ref.\cite{Barnett96},
implies the following $95\%$ CL
limit on the corresponding deviation from the Standard Model $\Delta
C_{1q}=\sum_{ij}(\sqrt{2}\pi \eta_{ij}\delta_{i})/(G_F \Lambda^{q2}_{ij} )
$ (where $\delta_i=+1,-1$ for $i=R$ or $i=L$, respectively):
\beq
-0.099<\Delta C_{1u}<0.051 ~~~~~~~
-0.050<\Delta C_{1d}<0.084
\eeq
Assuming a single non-vanishing operator at a time, this
translates into
\beq
\Lambda^{+u}_{LL},~\Lambda^{+u}_{LR},~\Lambda^{-u}_{RL},~\Lambda^{-u}_{RR}
>2.0~{\rm TeV},
\eeq
\beq
\Lambda^{-u}_{LL},~\Lambda^{-u}_{LR},~\Lambda^{+u}_{RL},~\Lambda^{+u}_{RR}
>2.7~{\rm TeV},
\eeq
\beq
\Lambda^{+d}_{LL},~\Lambda^{+d}_{LR},~\Lambda^{-d}_{RL},~\Lambda^{-d}_{RR}
>2.8~{\rm TeV},
\eeq
\beq
\Lambda^{-d}_{LL},~\Lambda^{-d}_{LR},~\Lambda^{+d}_{RL},~\Lambda^{+d}_{RR}
>2.1~{\rm TeV}.
\eeq
Comparing with
Fig.~\ref{fig:complim} we see that these limits are 
quite constraining on the individual
terms. But we also see
from Fig.~\ref{fig:complim} that we could, for example, take
$\Lambda^{+u}_{RL}=\Lambda^{+u}_{LR}$~\cite{added2}, which is parity
conserving, and add the corresponding contributions that are very similar
in shape. In this way the
limits are evaded and the fit is as good as for the separate
contributions.
\\[1cm]
\noindent {\Large \bf Appendix: Some Attempts to Understand the Possible
Effects of  Initial-State Radiation}\\[0.3cm] \noindent
In an attempt to gain more insight into the apparent spread 
in the invariant masses of the ZEUS events, and the fact
that their masses appear at first sight to be somewhat higher
than those of the H1 events, we have looked into the possible 
impact of initial-state radiation (ISR)~\footnote{It is clear that
this exercise is best carried out by the experimental collaborations
themselves: our intention is only to form an approximate
impression of how large the effects of ISR might be.}. In the presence
of ISR, the relation between the value $M_e$ of the reconstructed resonance
mass, as determined by the 
electron method, is related to the true value $M$ by
\begin{equation}                                
M_e^2 = M^2 {(1 - {z \over y_e}) \over (1 - z)^2}
\label{xe}                                  
\end{equation}
where $z$ is the fraction of the electron's longitudinal momentum
lost to ISR, and $y_e$ is the value of the conventional deep-inelastic
$y$ variable estimated using the
electron method. In the case of small $z$, eq.~(\ref{xe}) reduces to:
\begin{equation}
M_e^2 = M^2 \left[ 1 + z ~\left({2y_e-1 \over y_e}\right) \right] \; ,
\label{xe2}                                       
\end{equation}
which corresponds to a negative shift in mass for $y_e<1/2$, and to a positive
shift for $y_e>1/2$.                                                 

The analogous relation between $M_{2\alpha}$, the
mass determined by the double-angle method,
and the true value $M$ is
\begin{equation}     
M_{2\alpha}^2 = M^2 {1 \over (1 - z)^2}~.
\label{xda}  
\end{equation}
In the presence of ISR,
$M_{2\alpha}$ will therefore always be larger than the true value of $M$. In
particular, $M_{2\alpha}$ will always be larger than $M_e$.            
                                                       
We recall that H1 prefers to estimate $M$ using $M_e$~\cite{H1}, 
whereas ZEUS favours $M_{2\alpha}$~\cite{ZEUS}. ISR effects could
therefore lead qualitatively to $M_{ZEUS}$ 
being greater than                
$M_{H1}$, as observed, and resolution differences might explain their
greater spread. On the other hand, the experimental cuts allow for a 
fraction of electron 
energy lost to ISR up to $\sim 10 \%$, so it is not clear 
whether its effect could be important quantitatively. 

One can use eqs.~(\ref{xe}) and (\ref{xda}) to extract a relation between the
masses reconstructed with the two techniques and the true mass, under the
hypothetical assumption that the measured differences are due to ISR and
not to
resolution effects. The following relations hold:
\bea                              
       z &=& y_e (1-\rho) \quad , \quad
       \rho=\frac{M_e^2}{M_{2\alpha}^2}  \label{zvsy} \\
  \label{truem}
       M^2 &=& M_{2\alpha}^2 \left( 1-y_e+y_e\rho \right)^2 \; .
\eea
We applied these relations to the five ZEUS candidate events, using the values
of $y$ extracted using the double-angle techique. Four out of the five events
have $x_e<x_{2\alpha}$, and are therefore compatible with the ISR
interpretation. The values of $z$ and of the true mass which we calculate using
the above relations are given in Table~\ref{tab:isr}.
\begin{table}                                                                
\begin{center}
\begin{tabular}{|l|cccc|} \hline
Ev. \# & $z$ & $M_e$ & $M_{2\alpha}$ & $M$ \\
1     & $\sim 0$ & 217 & 208 & 208 \\    
2     & 0.029 & 220 & 226 & 220 \\
3     & 0.027 & 225 & 235 & 229 \\
4     & 0.10 & 233 & 253 & 226 \\
5     & 0.073 & 200 & 231 & 215 \\ \hline
\end{tabular}                                     
\ccaption{}{\label{tab:isr}
Estimated effect of ISR on the apparent
masses of the five candidate ZEUS~\cite{ZEUS} events extracted
using the {\em electron} ($M_e$) and the {\em double-angle} ($M_{2\alpha}$)
techniques, where
$z$ is the energy fraction lost to ISR as estimated from eq.~(\ref{zvsy})
and $M$ is the ISR-corrected mass defined in eq.~(\ref{truem}).
The first          
event gives no indication that $z > 0$, and we have retained
the $M_{2\alpha}$ estimate.}
\end{center}                                                                   
\end{table} 

Allowing either the highest- or lowest-mass event to be background, we find
good consistency with a single mass value around 220 GeV, 
though this still looks higher
than that quoted by H1~\cite{H1}\footnote{We have checked that, when applied
to the H1 data, the above ISR estimation procedure makes no significant
difference to their preferred mass value $M\sim 200$ GeV, which has an energy
scale error of about 5 GeV.}.

Needless to say, only an accurate analysis by ZEUS, properly accounting for the
effects of experimental resolution and their correlations between the two
techniques, will provide an accurate estimate of the ISR-induced
corrections. As remarked in ref.~\cite{ZEUS}, the 
differences between the values 
of $M_e$ and $M_{2\alpha}$ observed in the five candidate events are also not
inconsistent with the reported measurement uncertainties.

\end{document}